# System reduction-based approximate reanalysis method for statically indeterminate structures with high-rank modification

Wenxiong Li*, Suiyin Chen, Huan Huang*

College of Water Conservancy and Civil Engineering, South China Agricultural University, Guangzhou 510642, China

*Corresponding author. E-mail: leewenxiong@scau.edu.cn (W. Li), happyhuang@scau.edu.cn (H. Huang)

**Abstract** Efficient structural reanalysis for high-rank modification plays an important role in engineering computations which require repeated evaluations of structural responses, such as structural optimization and probabilistic analysis. To improve the efficiency of engineering computations, a novel approximate static reanalysis method based on system reduction and iterative solution is proposed for statically indeterminate structures with high-rank modification. In this approach, a statically indeterminate structure is divided into the basis system and the additional components. Subsequently, the structural equilibrium equations are rewritten as an equation system with the stiffness matrix of the basis system and the pseudo forces derived from the additional elements. With the introduction of the spectral decomposition, a reduced equation system with the element forces of the additional elements as unknowns is established. The approximate solutions of the modified structure can then be obtained by solving the reduced equation system through a pre-conditioned iterative solution algorithm. The computational costs of the proposed method are compared with those of two other reanalysis methods, and numerical examples including static reanalysis and static nonlinear analysis are presented. The results demonstrate that the proposed method has excellent computational performance for both structures with homogeneous material and structures composed of functionally graded beams. Furthermore, the superiority of the proposed method suggests that the combination of system reduction and pre-conditioned iterative solution technology is an effective approach to develop high-performance reanalysis methods.

**Keywords** Structural reanalysis; System reduction; Iterative solution; High-rank modification; Pseudo forces

## 1. Introduction

Structural reanalysis plays a crucial role in various branches of structural engineering, such as optimization of large-scale structures, structural damage identification, nonlinear structural analysis and probabilistic analysis [1]. These fields require repeated evaluations of structural responses and updates to structural parameters. Structural reanalysis methods refer to the analysis techniques that achieve high computational efficiency for modified structures by utilizing the structural stiffness matrix, the displacement solution of the initial structure and other information [2]. In many practical applications, it is essential to efficiently perform the structural reanalysis under the conditions of extensive or even global modification. Therefore, the development of efficient reanalysis methods for the structures with high-rank modification is a topic of great interest.

Over the past few decades, various structural reanalysis methods have been developed and they can be classified into direct methods and approximated methods. Direct methods aim to provide exact solutions for the modified structures and are mostly variations of two general matrix-update formulas: the Sherman-Morrison (SM) [3] and Woodbury (SMW) formulas [4]. Akgün [2] reviewed the development of SMW formulas and investigated other structural reanalysis methods based on SMW formulas. Direct methods for structural static reanalysis based on SMW formulas were developed by Krisch and Rubinstein [5] and Huang and Verchery [6]. Derived from SMW formulas, Deng and Ghosn [7] designed a pseudo force solver, while Jia et al. [8] proposed the inexact Newton Woodbury method, both to improve computational efficiency for nonlinear analysis. Recently, SMW formulas have been used to develop efficient dynamic reanalysis methods, including vibration analysis [9], frequency response function analysis [10] and nonlinear time-history analysis [11]. Besides, other studies provide new direct methods for structural reanalysis, such as Song et al.'s exact reanalysis approach [12], the exact block-based reanalysis method [13], the enhanced substructure coupling dynamic reanalysis technique [14], and the reanalysis algorithm based on updating matrix triangular factorization [15] and based on change threshold [16]. Notably, most exact reanalysis methods are only suitable for structures with low-rank modification. With regard to high-rank modification, most direct reanalysis methods struggle to significantly improve solution efficiency.

In contrast to the direct methods, approximate methods are generally applied to the reanalysis of structures with a large number of small-magnitude changes. From an implementation perspective, approximate methods can be divided into two categories: those based on the combination of basis vectors and those based on an iterative solution process. The combined approximations (CA) method [17-21] is the most representative method under the concept of combining basis vectors. However, there are two complications that affect its applicability: determining the appropriate number of basis vectors and ensuring solution accuracy in cases of local large modification. Krylov subspace methods with an iterative process [22], such as the Pre-conditioned Conjugate Gradient method (PCG), Generalized Minimum Residual algorithm (GMRES) and Bi-Conjugate Gradient Stabilized algorithm (Bi-CGSTAB), are commonly used to obtain high-precision approximate solutions. PCG is one of the most practical approximate methods based on an iterative solution process. In PCG, the original linear system is transformed into a related system with a smaller condition number by introducing an appropriate pre-conditioning process [23], which reduces the total number of iterations required for solving the system within specified tolerances. PCG has been developed for structural static reanalysis with unchanged degrees of freedom (DOFs) [24, 25], added DOFs [26, 27] and general layout modifications [28]. Recently, topology optimization has been efficiently carried out using a reanalysis technique based on PCG [29, 30] and novel PCG methods have been developed for complex systems [31, 32]. The Krylov subspace methods, such as PCG, can perform structural reanalysis with high-rank modification. Generally, the computational effort of Krylov subspace methods at each iteration step is related to the scale of the structural stiffness equations. When the total number of DOFs for an overall structure is large, the computational effort at each iteration step will also be substantial.

The integration of system reduction and the Krylov subspace methods has the potential to facilitate more efficient static reanalysis. By combining system reduction with iterative solution, Li and Chen [33] proposed an efficient reanalysis method for structures with local modifications. This approach involves reconstructing the reduced equation system of the modified structure based on the distribution of modified elements and employing a pre-conditioned iterative solution algorithm to solve the reduced equation system. Despite its ability to achieve excellent performance in reanalysis for local modifications, Li and Chen's method is limited in its capacity to significantly improve computational efficiency for structures with high-rank modification due to the substantial scale of the reduced system. To expedite reanalysis for structures with high-rank modification, it is necessary to develop a system reduction method for structures with global modification and subsequently apply it in approximate reanalysis methods. Recently, Yang [34] proposed an exact method for structural static reanalysis with high-rank modification based on the flexibility disassembly perturbation and the SMW formulas. Subsequently, Yang and Peng [35] further developed a fast calculation method for sensitivity analysis. The authors contend that their achievements [34, 35] facilitate the construction of a reduced system under high-rank modification. Consequently, an efficient structural reanalysis method that combines system reduction and iterative solution for high-rank modification is both feasible and highly anticipated in light of their work.

This paper presents the development of an approximate reanalysis method for statically indeterminate structures with high-rank modification based on system reduction. In this approach, a statically indeterminate structure is separated into the basis system (a statically determinate structure) and the additional components. The structural equilibrium equations are subsequently reformulated as an equation system comprising the stiffness matrix of the basis system and pseudo forces introduced to account for the influence of the additional elements. By incorporating spectral decomposition, a reduced equation system with the element forces of the additional elements as unknowns is derived. The approximate solutions of the modified structure can then be obtained by solving the reduced equation system using a pre-conditioned iterative solution algorithm. The computational costs of the proposed method are compared with those of two other reanalysis methods, and numerical examples are presented to demonstrate its effectiveness and efficiency.

## 2. Formulation

*2.1. Concept of the proposed reanalysis method*

**Fig. 1** presents a schematic diagram of the proposed reanalysis method, which illustrates the concept of the method using a statically indeterminate structure composed of seven truss elements. As depicted in **Fig. 1**, the structure is partitioned into a basis system comprising six elements and an additional component (element 7). In the proposed method, the statically indeterminate structure that requires solution is treated as a basic system under the influence of external loads and additional forces induced by the additional components. Based on the force-deformation relationship of the additional components and the relationship between element deformation and structural nodal displacements for all elements, a reduced equation system can be constructed in accordance with the consistency of deformation expressions for the additional components. The reduced equation system is subsequently solved by an iterative approach based on Krylov subspace methods, and structural displacements can be obtained based on the displacement expression of the basis system. Taking the structure in the figure as an example, the size of the established reduced equation system is 1, thereby obviating the need to solve a linear equation system with size 6 for the original statically indeterminate structure. Furthermore, combining system reduction with iterative solution facilitates efficient implementation of reanalysis for large-scale structures.

The formulation and implementation of the proposed reanalysis method will be elaborated upon in subsequent sections.

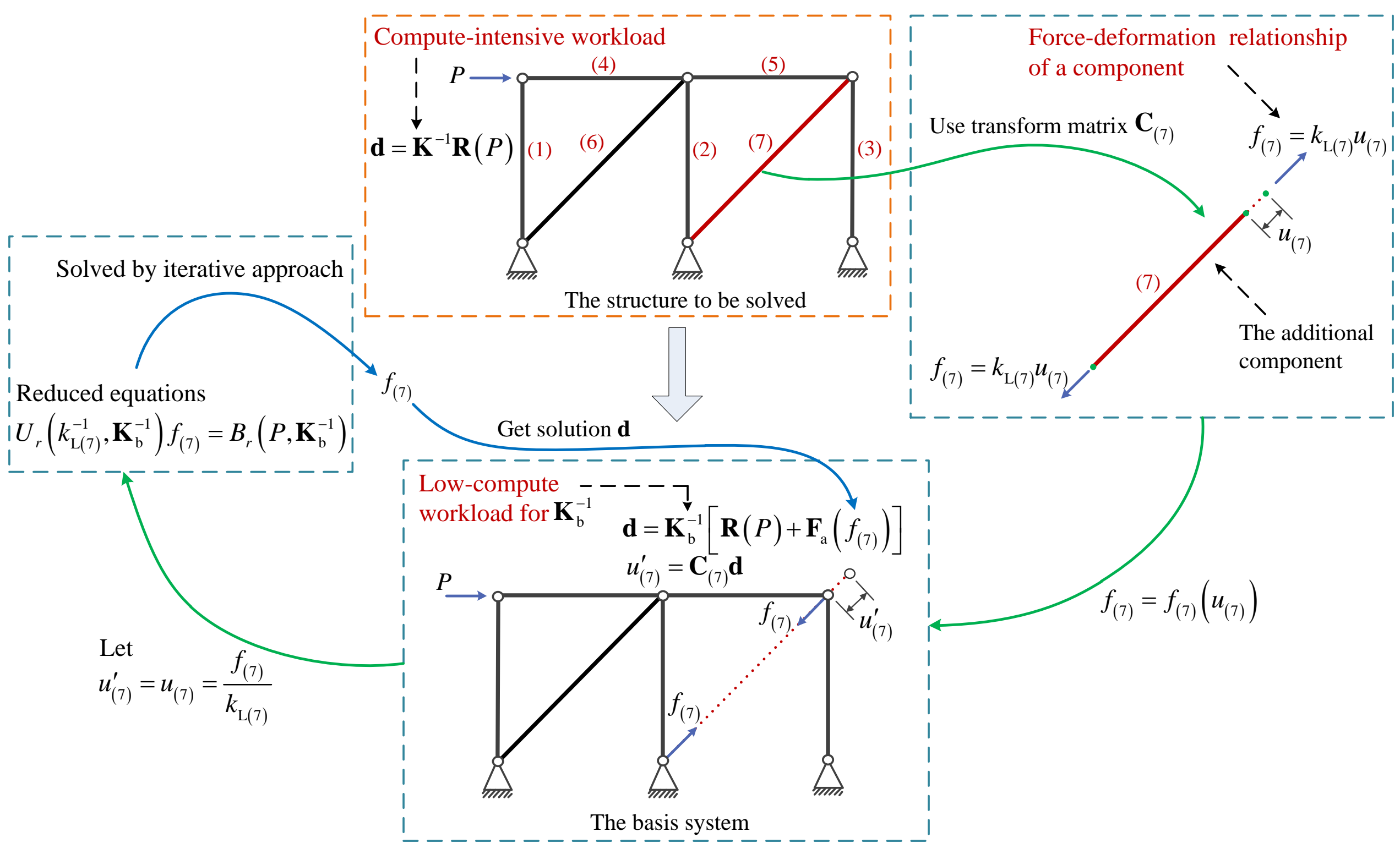

**Fig. 1**. The schematic diagram of the proposed reanalysis method.

### *2.2. Decomposition of stiffness matrices*

Regarding an $n_{(i)} \times n_{(i)}$ element stiffness matrix $\mathbf{k}_{(i)}$ with $i$ being the element index, let $\mathbf{X} \in \mathbb{C}^{n_{(i)}}$ be a nonzero vector for which

$$\mathbf{k}_{(i)}\mathbf{X} = \lambda \mathbf{X} \tag{1}$$

for some scalar $\lambda$. Then $\lambda$ is called an eigenvalue of the matrix $\mathbf{k}_{(i)}$ and $\mathbf{X}$ is called an eigenvector of $\mathbf{k}_{(i)}$ associated with $\lambda$. The $n_{(i)}$ eigenvalues $\lambda^{(j)}\left(j = 1, 2, \ldots, n_{(i)}\right)$ can be obtained by solving

$$\left|\mathbf{k}_{(i)} - \lambda \mathbf{I}\right| = 0 \tag{2}$$

where $\mathbf{I}$ is the identity matrix, and the eigenvectors $\mathbf{X}^{(j)}\left(j = 1, 2, \ldots, n_{(i)}\right)$ corresponding to $\lambda^{(j)}\left(j = 1, 2, \ldots, n_{(i)}\right)$ can be determined by finding the basic solutions to

$$\left(\mathbf{k}_{(i)} - \lambda^{(j)}\mathbf{I}\right)\mathbf{X}^{(j)} = 0 \tag{3}$$

Especially, the eigenvectors should be normalized to satisfy $\left[\mathbf{X}^{(j)}\right]^{\mathrm{T}}\left[\mathbf{X}^{(j)}\right] = 1$ and each eigenvector is orthogonal to one another. Note that the relation $\lambda^{(j)} = \left[\mathbf{X}^{(j)}\right]^{\mathrm{T}}\mathbf{k}_{(i)}\mathbf{X}^{(j)}$ holds and the matrix $\mathbf{P} = \left[\mathbf{X}^{(1)} \quad \mathbf{X}^{(2)} \quad \cdots \quad \mathbf{X}^{\left(n_{(i)}\right)}\right]$ is an orthogonal matrix.

In the implementation, the eigenvalues and eigenvectors of each element stiffness matrix within the structure can be readily obtained by utilizing the EIG(•) function provided by MATLAB.

An element stiffness matrix is always rank-deficient. This implies that for the $n_{(i)} \times n_{(i)}$ element stiffness matrix with rank $m_{(i)}\left(m_{(i)} < n_{(i)}\right)$, there are $n_{(i)} - m_{(i)}$ zero eigenvalues, which correspond to the rigid body translation and rigid body rotation modes of the element. For the plane elements, $m_{(i)} = n_{(i)} - 3$ because there are two rigid body translation modes and one rigid body rotation mode; for the spatial elements, $m_{(i)} = n_{(i)} - 6$ because there are three rigid body translation modes and three rigid body rotation modes. Then, the $m_{(i)}$ nonzero eigenvalues $\lambda^{(j)}\left(\lambda^{(j)} \neq 0, j = 1, 2, \ldots, m_{(i)}\right)$ can be regarded as the stiffness parameters associated with the independent deformation modes.

For an element which has $m_{(i)}$ independent deformation modes, the element stiffness matrix $\mathbf{k}_{(i)}$ can be expressed as follows according to the decomposition provided in Ref. [34, 36, 37]

$$\mathbf{k}_{(i)} = \mathbf{c}_{(i)}^{\mathrm{T}}\mathbf{k}_{\mathrm{L}(i)}\mathbf{c}_{(i)} \tag{4}$$

where $\mathbf{k}_{\mathrm{L}(i)} = \mathrm{diag}\left(\lambda^{(1)}, \lambda^{(2)}, \ldots, \lambda^{\left(m_{(i)}\right)}\right)$ indicates the $m_{(i)} \times m_{(i)}$ matrix composed of stiffness parameters of the $i$-th element, with $m_{(i)}$ as the rank of $\mathbf{k}_{\mathrm{L}(i)}$; $\mathbf{c}_{(i)} = \left[\mathbf{X}^{(1)} \quad \mathbf{X}^{(2)} \quad \cdots \quad \mathbf{X}^{\left(m_{(i)}\right)}\right]^{\mathrm{T}}$ designates a transform matrix between the element stiffness matrix and the elemental stiffness parameters, which is composed of the $m_{(i)}$ eigenvectors corresponding to the nonzero eigenvalues.

The size of $\mathbf{c}_{(i)}$ is $m_{(i)} \times n_{(i)}$ and each line of $\mathbf{c}_{(i)}$ reflects the topological connectivity between DOFs for the corresponding elemental stiffness parameters in $\mathbf{k}_{\mathrm{L}(i)}$. With respect to the structure with a given topology, $\mathbf{c}_{(i)}$ is considered to be invariant. Eq. (4) is called spectral decomposition in this paper. For more details of the spectral decomposition, readers are referred to Ref. [34, 36, 37].

Regarding a structure composed of $m$ elements, the stiffness matrix $\mathbf{K}$ is expressed by the form of the spectral decomposition as

$$\mathbf{K} = \sum_{i=1}^{m} \mathbf{k}_{(i)} = \sum_{i=1}^{m} \mathbf{c}_{(i)}^{\mathrm{T}} \mathbf{k}_{\mathrm{L}(i)} \mathbf{c}_{(i)} = \mathbf{C}^{\mathrm{T}} \mathbf{K}_{\mathrm{L}} \mathbf{C} \tag{5}$$

where the summation symbol $\sum_{i=1}^{m}(\cdot)$ denotes the conventional assembly of $m$ elements, $\mathbf{K}$ represents an $n \times n$ matrix with $n$ as the total number of DOFs of the structure, and $\mathbf{K}_{\mathrm{L}}$ refers to the matrix of structural stiffness parameters, which is a block diagonal matrix. The non-zero components of $\mathbf{K}_{\mathrm{L}}$ are composed of the stiffness parameter matrices of all elements, namely,

$$\mathbf{K}_{\mathrm{L}} = \sum_{i=1}^{m} \mathbf{k}_{\mathrm{L}(i)} = \begin{bmatrix} \mathbf{k}_{\mathrm{L}(1)} & \mathbf{0} & \cdots & \mathbf{0} \\ \mathbf{0} & \mathbf{k}_{\mathrm{L}(2)} & \cdots & \mathbf{0} \\ \vdots & \vdots & \ddots & \vdots \\ \mathbf{0} & \mathbf{0} & \cdots & \mathbf{k}_{\mathrm{L}(m)} \end{bmatrix} \tag{6}$$

In Eq. (5), $\mathbf{C}$ represents the transform matrix between the matrix of structural stiffness parameters and the stiffness matrix of the whole structure, expressed as

$$\mathbf{C} = \sum_{i=1}^{m} \mathbf{c}_{(i)} = \begin{bmatrix} \mathbf{C}_{(1)}^{\mathrm{T}} & \mathbf{C}_{(2)}^{\mathrm{T}} & \cdots & \mathbf{C}_{(m)}^{\mathrm{T}} \end{bmatrix}^{\mathrm{T}} \tag{7}$$

Notably, $\mathbf{C}_{(i)}\left(i = 1, 2, \cdots, m\right)$ denotes the extended matrix of $\mathbf{c}_{(i)}\left(i = 1, 2, \cdots, m\right)$ for reflecting the topological connectivity between DOFs of the whole structure for the corresponding elemental stiffness parameters in $\mathbf{k}_{\mathrm{L}(i)}$, and the size of $\mathbf{C}_{(i)}\left(i = 1, 2, \cdots, m\right)$ is $m_{(i)} \times n$. $\mathbf{C}_{(i)}\left(i = 1, 2, \cdots, m\right)$ is also invariant for the structure with given topology.

*2.3. Reduced system of statically indeterminate structures*

With respect to a structure with $n$ DOFs, the equation system is expressed as

$$\mathbf{Kd} = \mathbf{R} \tag{8}$$

where $\mathbf{R}$ and $\mathbf{d}$ are the external load vector and the structural nodal displacement vector, respectively, and their size is $n \times 1$. The statically indeterminate structure is separated into two parts: the basis system and additional components. Generally, the basis system is a statically determinate structure, while the additional components consist of the components other than the basis system. Particularly, the additional components are distributed in the structure and are not required to form a complete structural system. Corresponding to the basis system and the additional components, the stiffness matrix of the structure is separated into two parts. Then, Eq. (8) is further expressed as

$$\left(\mathbf{K}_{\mathrm{b}} + \mathbf{K}_{\mathrm{a}}\right)\mathbf{d} = \mathbf{R} \tag{9}$$

where $\mathbf{K}_{\mathrm{b}}$ and $\mathbf{K}_{\mathrm{a}}$ denote the stiffness matrices of the basis system and the additional components, respectively, and the size of both is $n \times n$. For the system without additional components, Eq. (9) can be written as $\mathbf{K}_{\mathrm{b}}\mathbf{d} = \mathbf{R}$. When subjected to external loads, the additional components within the structure will undergo deformation and generate internal forces. Utilizing a concept similar to that presented in Ref. [38], the nodal forces produced by these additional components are considered as pseudo forces acting upon the basis system. Then, Eq. (9) is further expressed as

$$\mathbf{K}_{\mathrm{b}}\mathbf{d} = \mathbf{R} - \mathbf{K}_{\mathrm{a}}\mathbf{d} \tag{10}$$

where $\mathbf{K}_{\mathrm{a}}\mathbf{d}$ represents the pseudo nodal forces generated by the elements corresponding to the additional components. The expression of the nodal displacement vector can be rewritten as

$$\mathbf{d} = \mathbf{K}_{\mathrm{b}}^{-1}\mathbf{R} - \mathbf{K}_{\mathrm{b}}^{-1}\mathbf{K}_{\mathrm{a}}\mathbf{d} \tag{11}$$

With the expression in Eq. (5), $\mathbf{K}_{\mathrm{b}}$ and $\mathbf{K}_{\mathrm{a}}$ are expressed as

$$\mathbf{K}_{\mathrm{b}} = \sum_{i \in \mathrm{BASESYS}} \mathbf{k}_{(i)} = \sum_{i \in \mathrm{BASESYS}} \mathbf{c}_{(i)}^{\mathrm{T}}\mathbf{k}_{\mathrm{L}(i)}\mathbf{c}_{(i)} = \mathbf{C}_{\mathrm{b}}^{\mathrm{T}}\mathbf{K}_{\mathrm{Lb}}\mathbf{C}_{\mathrm{b}} \tag{12}$$

$$\mathbf{K}_{\mathrm{a}} = \sum_{i \in \mathrm{ADDSYS}} \mathbf{k}_{(i)} = \sum_{i \in \mathrm{ADDSYS}} \mathbf{c}_{(i)}^{\mathrm{T}}\mathbf{k}_{\mathrm{L}(i)}\mathbf{c}_{(i)} = \mathbf{C}_{\mathrm{a}}^{\mathrm{T}}\mathbf{K}_{\mathrm{La}}\mathbf{C}_{\mathrm{a}} \tag{13}$$

where BASESYS and ADDSYS represent the sets of elements for the basis system and the additional components, respectively; $\mathbf{K}_{\mathrm{Lb}}$ and $\mathbf{K}_{\mathrm{La}}$ indicate the matrices of stiffness parameters for the basis system and the additional components, respectively. In Eq. (12), $\mathbf{C}_{\mathrm{b}}$ and $\mathbf{K}_{\mathrm{Lb}}$ are both full rank matrices with the size of $n \times n$. Therefore, the inverse of $\mathbf{K}_{\mathrm{b}}$ is obtained by

$$\mathbf{K}_{\mathrm{b}}^{-1} = \mathbf{C}_{\mathrm{b}}^{-1} \mathbf{K}_{\mathrm{Lb}}^{-1} \mathbf{C}_{\mathrm{b}}^{-\mathrm{T}} \tag{14}$$

By substituting Eqs. (14) and (13) into Eq. (11), the nodal displacement vector is expressed as

$$\mathbf{d} = \mathbf{C}_{\mathrm{b}}^{-1} \mathbf{K}_{\mathrm{Lb}}^{-1} \mathbf{C}_{\mathrm{b}}^{-\mathrm{T}} \mathbf{R} - \mathbf{C}_{\mathrm{b}}^{-1} \mathbf{K}_{\mathrm{Lb}}^{-1} \mathbf{C}_{\mathrm{b}}^{-\mathrm{T}} \mathbf{K}_{\mathrm{a}} \mathbf{d} \tag{15}$$

In Eq. (15), $\mathbf{K}_{\mathrm{a}}\mathbf{d}$ can be further expressed as

$$\mathbf{K}_{\mathrm{a}} \mathbf{d} = \sum_{e \in \mathrm{ADDSYS}} \mathbf{c}_{(e)}^{\mathrm{T}} \mathbf{k}_{\mathrm{L}(e)} \mathbf{u}_{(e)} = \mathbf{C}_{\mathrm{a}}^{\mathrm{T}} \mathbf{K}_{\mathrm{La}} \mathbf{u}_{\mathrm{a}} \tag{16}$$

where $\mathbf{u}_{(e)}$ represents the generalized deformation vector of the $e$-th element, corresponding to the stiffness parameters of the $e$-th element, and $\mathbf{u}_{\mathrm{a}}$ is expressed as

$$\mathbf{u}_{\mathrm{a}} = \begin{bmatrix} \mathbf{u}_{(1)}^{\mathrm{T}} & \mathbf{u}_{(1)}^{\mathrm{T}} & \cdots & \mathbf{u}_{(m)}^{\mathrm{T}} \end{bmatrix}^{\mathrm{T}} \tag{17}$$

where $m$ refers to the number of elements corresponding to the additional components. By substituting Eq. (16) into Eq. (15), the nodal displacement vector is expressed by the generalized deformation vector of the additional components as

$$\mathbf{d} = \mathbf{C}_{\mathrm{b}}^{-1} \mathbf{K}_{\mathrm{Lb}}^{-1} \mathbf{C}_{\mathrm{b}}^{-\mathrm{T}} \mathbf{R} - \mathbf{C}_{\mathrm{b}}^{-1} \mathbf{K}_{\mathrm{Lb}}^{-1} \mathbf{C}_{\mathrm{b}}^{-\mathrm{T}} \mathbf{C}_{\mathrm{a}}^{\mathrm{T}} \mathbf{K}_{\mathrm{La}} \mathbf{u}_{\mathrm{a}} \tag{18}$$

Considering the relationship between the generalized deformation vector of an element and the structural nodal displacement vector:

$$\mathbf{u}_{(i)} = \mathbf{C}_{(i)} \mathbf{d}, \tag{19}$$

the expression of the generalized deformation vector of the additional components is obtained by substituting Eq. (18) into Eq. (19) and using the relationship in Eq. (17):

$$\mathbf{u}_{\mathrm{a}} = \mathbf{C}_{\mathrm{s}} \mathbf{K}_{\mathrm{Lb}}^{-1} \mathbf{C}_{\mathrm{b}}^{-\mathrm{T}} \mathbf{R} - \mathbf{C}_{\mathrm{s}} \mathbf{K}_{\mathrm{Lb}}^{-1} \mathbf{C}_{\mathrm{s}}^{\mathrm{T}} \mathbf{K}_{\mathrm{La}} \mathbf{u}_{\mathrm{a}} \tag{20}$$

where

$$\mathbf{C}_{\mathrm{s}} = \mathbf{C}_{\mathrm{a}}\mathbf{C}_{\mathrm{b}}^{-1} \tag{21}$$

By defining the generalized force vector corresponding to the generalized deformation vector of the additional components as

$$\mathbf{F}_{\mathrm{a}} = \mathbf{K}_{\mathrm{La}}\mathbf{u}_{\mathrm{a}}, \tag{22}$$

the generalized deformation vector $\mathbf{u}_{\mathrm{a}}$ can be expressed by $\mathbf{F}_{\mathrm{a}}$ as

$$\mathbf{u}_{\mathrm{a}} = \mathbf{K}_{\mathrm{La}}^{-1}\mathbf{F}_{\mathrm{a}} \tag{23}$$

By substituting Eq. (23) into Eq. (20) , the new equation system with the generalized force vector of the additional components as unknowns can be established as

$$\mathbf{U}_{\mathrm{r}}\left(\mathbf{K}_{\mathrm{La}}^{-1}, \mathbf{K}_{\mathrm{Lb}}^{-1}\right)\mathbf{F}_{\mathrm{a}} = \mathbf{B}_{\mathrm{r}}\left(\mathbf{R}, \mathbf{K}_{\mathrm{Lb}}^{-1}\right) \tag{24}$$

where

$$\begin{aligned} \mathbf{U}_{\mathrm{r}}\left(\mathbf{K}_{\mathrm{La}}^{-1}, \mathbf{K}_{\mathrm{Lb}}^{-1}\right) &= \mathbf{K}_{\mathrm{La}}^{-1} + \mathbf{C}_{\mathrm{s}}\mathbf{K}_{\mathrm{Lb}}^{-1}\mathbf{C}_{\mathrm{s}}^{\mathrm{T}} \\ \mathbf{B}_{\mathrm{r}}\left(\mathbf{R}, \mathbf{K}_{\mathrm{Lb}}^{-1}\right) &= \mathbf{C}_{\mathrm{s}}\mathbf{K}_{\mathrm{Lb}}^{-1}\mathbf{C}_{\mathrm{b}}^{-\mathrm{T}}\mathbf{R} \end{aligned} \tag{25}$$

The size of the above equation system is $m_{\mathrm{a}} = \sum_{e \in \mathrm{ADDSYS}} m_{(e)}$ , which is associated with the elements of the additional components. In the case of $m_{\mathrm{a}} \le n$, which applies to most engineering structures, the computational workload is significantly reduced by solving the new equation system (Eq.(24)) instead of the original equation system (Eq. (8)). Besides, only matrix $\mathbf{K}_{\mathrm{La}}^{-1}$ and matrix $\mathbf{K}_{\mathrm{Lb}}^{-1}$ in Eq. (24) need to be updated for any modified structure. Note that $\mathbf{K}_{\mathrm{La}}$ and $\mathbf{K}_{\mathrm{Lb}}$ are both block diagonal matrices, and $\mathbf{K}_{\mathrm{La}}^{-1}$ and $\mathbf{K}_{\mathrm{Lb}}^{-1}$ are easily obtained by taking the inverse of each diagonal block of $\mathbf{K}_{\mathrm{La}}$ and $\mathbf{K}_{\mathrm{Lb}}$. In other words, the coefficient matrix and the right side of Eq. (24) are easily obtained without the time-consuming matrix inversion. It is worth mentioning that Eq. (24) can be easily solved by using a pre-conditioned iterative solution algorithm because the coefficient matrix of Eq. (24) is positive definite and symmetric.

To elucidate the mechanical significance of reduction system, Eq. (24) can be reformulated as $\mathbf{u}_{\mathrm{a}} = \mathbf{K}_{\mathrm{La}}^{-1}\mathbf{F}_{\mathrm{a}} = \mathbf{C}_{\mathrm{s}}\mathbf{K}_{\mathrm{Lb}}^{-1}\mathbf{C}_{\mathrm{b}}^{-\mathrm{T}}\mathbf{R} - \mathbf{C}_{\mathrm{s}}\mathbf{K}_{\mathrm{Lb}}^{-1}\mathbf{C}_{\mathrm{s}}^{\mathrm{T}}\mathbf{F}_{\mathrm{a}}$ . This reflects the deformation compatibility condition imposed upon the additional components, whereby the deformation of these components generated by the modified structure under external loads is consistent with the deformation of the additional components generated by the basis system under the action of both external loads and the generalized forces of the additional components.

*2.4. Iterative solution for the reduced system*

The Pre-conditioned Conjugate Gradient algorithm (PCG) is an iterative method employed to solve linear systems characterized by a positive definite and symmetric coefficient matrix. Its high computational efficiency has been empirically validated. By drawing upon the algorithmic framework of PCG, this paper constructs an iterative solution algorithm for the reduction system.

Generally, the right side of Eq. (24) is determined in advance and Eq. (24) is denoted as

$$\left(\mathbf{K}_{\mathrm{La}}^{-1}+\mathbf{C}_{\mathrm{s}}\mathbf{K}_{\mathrm{Lb}}^{-1}\mathbf{C}_{\mathrm{s}}^{\mathrm{T}}\right)\mathbf{x}=\mathbf{B} \tag{26}$$

where $\mathbf{B}=\mathbf{C}_{\mathrm{s}}\mathbf{K}_{\mathrm{Lb}}^{-1}\mathbf{C}_{\mathrm{b}}^{-\mathrm{T}}\mathbf{R}$ and $\mathbf{x}=\mathbf{F}_{\mathrm{a}}$. Regarding the pre-conditioning, the pre-conditioner $\mathbf{M}$ is set according to the initial structure, namely,

$$\mathbf{M}=\mathbf{K}_{\mathrm{La,0}}^{-1}+\mathbf{C}_{\mathrm{s}}\mathbf{K}_{\mathrm{Lb,0}}^{-1}\mathbf{C}_{\mathrm{s}}^{\mathrm{T}}, \tag{27}$$

where $\mathbf{K}_{\mathrm{Lb,0}}$ and $\mathbf{K}_{\mathrm{La,0}}$ represent the initial matrices of stiffness parameters for the basis system and the additional components, respectively. During the implementation of pre-conditioned iterative solution algorithm, $\mathbf{M}^{-1}$ is obtained in advance and applied to the subsequent iterative solution.

The implementation of iterative solution for solving Eq. (26) is shown in **Fig. 2**. In **Fig. 2**, $\mathbf{x}_j$, $\mathbf{r}_j$, $\mathbf{z}_j$ and $\mathbf{p}_j$ represent the solution, the residual vector, the pre-conditioned residual vector and the search direction vector at the $j$ iteration, respectively, and the operator $\left(\mathbf{v}_a,\mathbf{v}_b\right)$ represents the vector inner product of $\mathbf{v}_a$ and $\mathbf{v}_b$. The convergence criterion is set as $\left\|\mathbf{r}_j\right\|/\left\|\mathbf{B}\right\|<\varepsilon$, where $\varepsilon$ represents the tolerance. Once the convergence criterion is satisfied, the solution of the reduced system can be obtained as $\mathbf{F}_{\mathrm{a}}=\mathbf{x}_{j+1}$, and then the nodal displacement vector of the structure is obtained by

$$\mathbf{d} = \mathbf{C}_{\mathrm{b}}^{-1}\left(\mathbf{K}_{\mathrm{Lb}}^{-1}\mathbf{C}_{\mathrm{b}}^{-\mathrm{T}}\mathbf{R} - \mathbf{K}_{\mathrm{Lb}}^{-1}\mathbf{C}_{\mathrm{s}}^{\mathrm{T}}\mathbf{F}_{\mathrm{a}}\right), \tag{28}$$

according to Eq. (23) and Eq. (18).

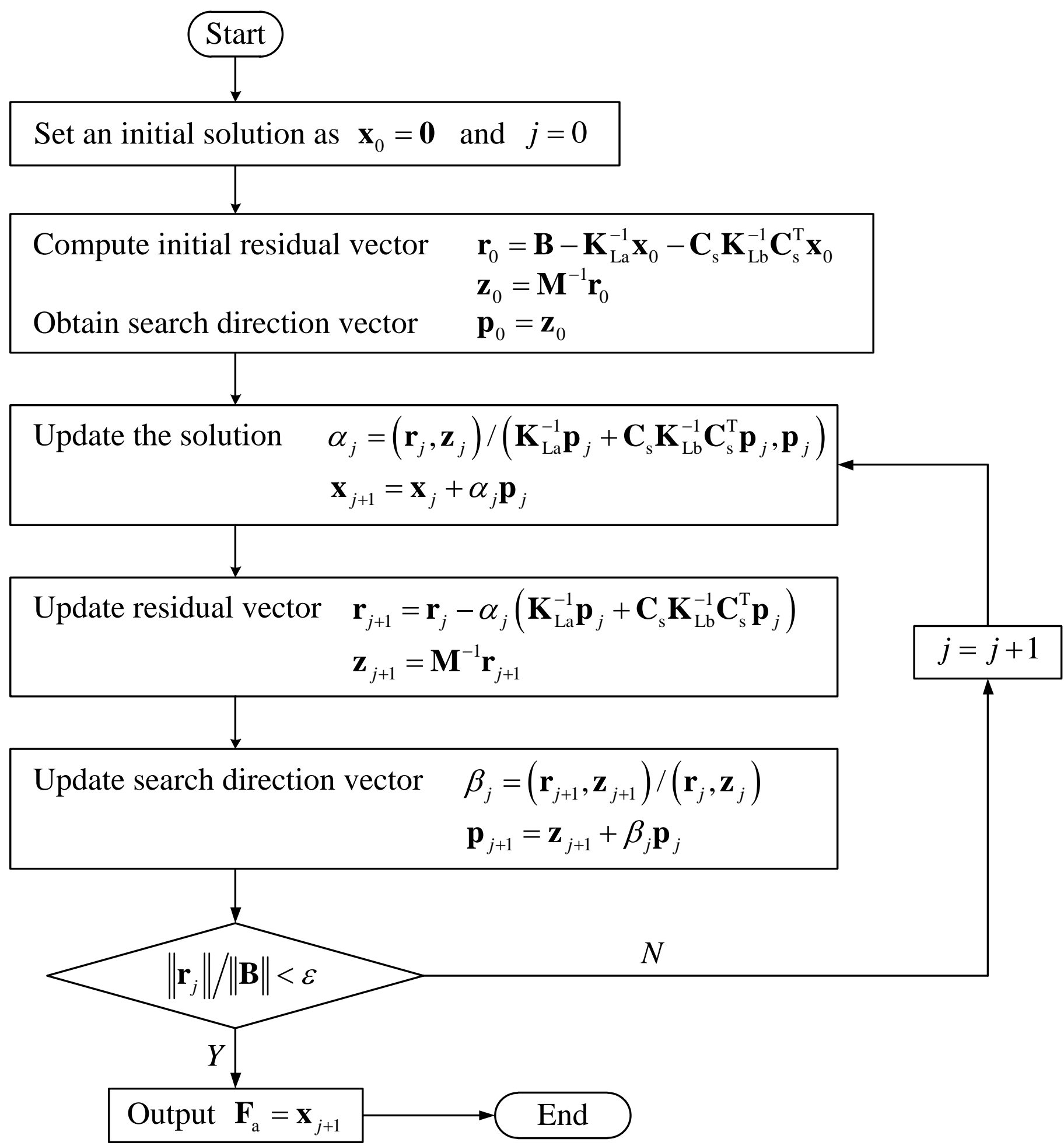

**Fig. 2**. Flowchart of iterative solution for the reduced system.

## 3. Computational cost

This section examines the computational cost of the proposed reanalysis method and compares it with two other existing reanalysis methods that are also applicable to high-rank modification. For ease of reference, the three reanalysis methods under consideration are denoted as follows: (1) SRI, the proposed structural reanalysis method based on System Reduction and Iterative solution; (2) PCG, an existing structural reanalysis method based on the Pre-conditioning Conjugate Gradient method [26]; and (3) FDP, the structural reanalysis method based on Flexibility Disassembly Perturbation [34]. Given that modern computing devices typically offer ample storage capacity, this paper does not consider the space complexity of computational methods. Instead, computational cost is quantified in terms of the number of floating-point operations (flops).

In the ensuing analysis, the number of DOFs for the statically indeterminate structure is denoted as $n$. The scale of the stiffness parameter matrix $\mathbf{K}_{\mathrm{La}}$ for the additional components is represented by $q$. To simplify the analysis, this section considers only the case in which the stiffness parameters for each element are uncoupled, such that both $\mathbf{K}_{\mathrm{La}}$ and $\mathbf{K}_{\mathrm{Lb}}$ are diagonal matrices. In instances where stiffness parameters are coupled, $\mathbf{K}_{\mathrm{La}}$ and $\mathbf{K}_{\mathrm{Lb}}$ may be block diagonal matrices, resulting in a slight increase in computational cost. However, this impact is negligible when compared to overall computational cost. The number of iterations required to satisfy the convergence criterion is denoted as $k$ for reanalysis methods that employ an iterative solution.

*3.1. Flops of SRI*

The implementation of reanalysis using SRI can be divided into three stages: (A) Obtaining the initial vector; (B) Iteratively solving the reduced system; and (C) Calculating the nodal displacement vector of the modified structure. Certain matrices and vectors used in SRI, such as $\mathbf{C}_{\mathrm{s}}$, $\mathbf{C}_{\mathrm{b}}$, $\mathbf{R}$ and $\mathbf{M}^{-1}$, can be determined prior to the reanalysis operation. For tasks involving multiple reanalyses, these matrices remain unchanged and the computational cost of obtaining them is relatively small compared to that of multiple reanalyses. As such, the calculation required to obtain these matrices is not included in the computational cost. **Table 1** presents the number of flops for each formula used in SRI.

**Table 1** Number of flops for the formulas in SRI.

| Stage | Index | Formula | Number of flops |
|---|---|---|---|
| A | A-1 | $\mathbf{B}=\mathbf{C}_{\mathrm{s}}\mathbf{K}_{\mathrm{Lb}}^{-1}\mathbf{C}_{\mathrm{b}}^{-\mathrm{T}}\mathbf{R}$ | $2n^2+2nq+n$ |
| | A-2 | $\mathbf{r}_0=\mathbf{B}-\left(\mathbf{K}_{\mathrm{La}}^{-1}\mathbf{x}_0+\mathbf{C}_{\mathrm{s}}\mathbf{K}_{\mathrm{Lb}}^{-1}\mathbf{C}_{\mathrm{s}}^{\mathrm{T}}\mathbf{x}_0\right)$ | $2q^2+4nq+2q+n$ |
| | A-3 | $\mathbf{z}_0=\mathbf{M}^{-1}\mathbf{r}_0$ | $2nq$ |
| B | B-1 | $\alpha_j=\left(\mathbf{r}_j,\mathbf{z}_j\right)/\left(\mathbf{K}_{\mathrm{La}}^{-1}\mathbf{p}_j+\mathbf{C}_{\mathrm{s}}\mathbf{K}_{\mathrm{Lb}}^{-1}\mathbf{C}_{\mathrm{s}}^{\mathrm{T}}\mathbf{p}_j,\mathbf{p}_j\right)$ | $2q^2+4nq+5q+n+1$ |
| | B-2 | $\mathbf{x}_{j+1}=\mathbf{x}_j+\alpha_j\mathbf{p}_j$ | $2q$ |
| | B-3 | $\mathbf{r}_{j+1}=\mathbf{r}_j-\alpha_j\left(\mathbf{K}_{\mathrm{La}}^{-1}\mathbf{p}_j+\mathbf{C}_{\mathrm{s}}\mathbf{K}_{\mathrm{Lb}}^{-1}\mathbf{C}_{\mathrm{s}}^{\mathrm{T}}\mathbf{p}_j\right)$ | $2q^2+4nq+3q+n$ |
| | B-4 | $\mathbf{z}_{j+1}=\mathbf{M}^{-1}\mathbf{r}_{j+1}$ | $2q^2$ |
| | B-5 | $\beta_j=\left(\mathbf{r}_{j+1},\mathbf{z}_{j+1}\right)/\left(\mathbf{r}_j,\mathbf{z}_j\right)$ | $4q+1$ |
| | B-6 | $\mathbf{p}_{j+1}=\mathbf{z}_{j+1}+\beta_j\mathbf{p}_j$ | $2q$ |
| C | C-1 | $\mathbf{d}=\mathbf{C}_{\mathrm{b}}^{-1}\left(\mathbf{K}_{\mathrm{Lb}}^{-1}\mathbf{C}_{\mathrm{b}}^{-\mathrm{T}}\mathbf{R}-\mathbf{K}_{\mathrm{Lb}}^{-1}\mathbf{C}_{\mathrm{s}}^{\mathrm{T}}\mathbf{F}_{\mathrm{a}}\right)$ | $2q^2+2nq+2n$ |

It is important to note that the majority of operations in SRI are performed through matrix-vector multiplication to minimize the number of flops, and no matrix multiplication is required during the solution process. For instance, the specific calculation order and number of flops for formula (A-1) are as follows

$$\mathbf{B}=\underbrace{\mathbf{C}_{\mathrm{s}}\underbrace{\mathbf{K}_{\mathrm{Lb}}^{-1}\underbrace{\mathbf{C}_{\mathrm{b}}^{-\mathrm{T}}\mathbf{R}}_{2n^2}}_{2n^2+n}}_{2n^2+n+2nq} \tag{29}$$

In Eq. (29), the result of $\mathbf{K}_{\mathrm{Lb}}^{-1}\mathbf{C}_{\mathrm{b}}^{-\mathrm{T}}\mathbf{R}$ can be saved and denoted as $\mathbf{B}_{\mathrm{s}}$ for the implementation of formula (C-1) in **Table 1**. Then, the specific calculation order and number of flops for formula (C-1) are

$$\mathbf{d}=\underbrace{\mathbf{C}_{\mathrm{b}}^{-1}\left(\underbrace{\mathbf{B}_{\mathrm{s}}-\overbrace{\mathbf{K}_{\mathrm{Lb}}^{-1}\overbrace{\mathbf{C}_{\mathrm{s}}^{\mathrm{T}}\mathbf{F}_{\mathrm{a}}}^{2nq}}^{2nq+n}}_{2nq+2n}\right)}_{2n^2+2nq+2n} \tag{30}$$

In addition, the calculation order and number of flops for $\mathbf{K}_{\mathrm{La}}^{-1}\mathbf{x}_0+\mathbf{C}_{\mathrm{s}}\mathbf{K}_{\mathrm{Lb}}^{-1}\mathbf{C}_{\mathrm{s}}^{\mathrm{T}}\mathbf{x}_0$ in formula (A-2) are

$$\underbrace{\underbrace{\mathbf{K}_{\mathrm{La}}^{-1}\mathbf{x}_0}_{2q^2} + \underbrace{\mathbf{C}_{\mathrm{s}}\, \underbrace{\mathbf{K}_{\mathrm{Lb}}^{-1}\, \underbrace{\mathbf{C}_{\mathrm{s}}^{\mathrm{T}}\mathbf{x}_0}_{2nq}}_{2nq+n}}_{4nq+n}}_{2q^2+4nq+n} \tag{31}$$

Similarly, the number of flops of $\mathbf{K}_{\mathrm{La}}^{-1}\mathbf{p}_j + \mathbf{C}_{\mathrm{s}}\mathbf{K}_{\mathrm{Lb}}^{-1}\mathbf{C}_{\mathrm{s}}^{\mathrm{T}}\mathbf{p}_j$ in formulas (B-1) and (B-3) is $2q^2 + 4nq + n$. The number of flops for vector inner product, such as $\left(\mathbf{r}_j, \mathbf{z}_j\right)$ in formulas (B-1) and (B-5), is $2q$.

According to the statistics of the number of flops in **Table 1**, the total number of flops for SRI with *k* iterations is

$$T_{F-SRI} = k\left(6q^2 + 8nq + 2n + 16q + 2\right) + 2n^2 + 4q^2 + 10nq + 4n + 2q \tag{32}$$

*3.2. Flops of PCG and FDP*

In contrast to the proposed SRI, PCG [22] directly solve Eq. (5) iteratively, where the scale of the equation system is $n$. In PCG, the initial structural stiffness matrix is treated as the pre-conditioner ($\mathbf{M} = \mathbf{K}_0$), and its inverse can be obtained and stored in advance. Similar to SRI, the calculation required to obtain $\mathbf{M}^{-1}$ is not included in computational cost. The iteration process of PCG is similar to that of SRI, with the matrix $\mathbf{K}_{\mathrm{La}}^{-1} + \mathbf{C}_{\mathrm{s}}\mathbf{K}_{\mathrm{Lb}}^{-1}\mathbf{C}_{\mathrm{s}}^{\mathrm{T}}$ being replaced by $\mathbf{K}$. The implementation of reanalysis using PCG can be divided into two stages: (A) Obtaining the initial vector; (B) Iteratively solving the system. Statistics similar to those for SRI are generated, and the total number of flops for PCG with *k* iterations is as follows

$$T_{F-PCG} = k\left(6n^2 + 14n + 2\right) + 4n^2 + n \tag{33}$$

FDP [34] is another reanalysis method for structures with high-rank modification. This method reduces computational effort by distinguishing the basic system and additional components and using the spectral decomposition. This section analyzes the computational cost of FDP based on the expressions provided in FDP and highlights the difference between FDP and SRI.

Firstly, the stiffness matrix of the modified structure is divided into

$$\mathbf{K} = \mathbf{K}_{\mathrm{b}} + \mathbf{K}_{\mathrm{a}} = \mathbf{K}_{\mathrm{b}} + \mathbf{C}_{\mathrm{a}}^{\mathrm{T}}\mathbf{K}_{\mathrm{La}}\mathbf{C}_{\mathrm{a}} \tag{34}$$

Then, according to the SMW formulas used in FDP, the inverse of the stiffness matrix is

$$\mathbf{K}^{-1}=\mathbf{K}_{\mathrm{b}}^{-1}-\mathbf{K}_{\mathrm{b}}^{-1}\mathbf{C}_{\mathrm{a}}^{\mathrm{T}}\mathbf{K}_{\mathrm{La}}\left(\mathbf{I}+\mathbf{C}_{\mathrm{a}}\mathbf{K}_{\mathrm{b}}^{-1}\mathbf{C}_{\mathrm{a}}^{\mathrm{T}}\mathbf{K}_{\mathrm{La}}\right)^{-1}\mathbf{C}_{\mathrm{a}}\mathbf{K}_{\mathrm{b}}^{-1} \tag{35}$$

By substituting Eqs. (14) and (21) into Eq. (35), the expression of nodal displacement vector of the modified structure can be obtained by using Eq. (8) and expressed as

$$\mathbf{d}=\mathbf{K}^{-1}\mathbf{R}=\mathbf{C}_{\mathrm{b}}^{-1}\left[\mathbf{K}_{\mathrm{Lb}}^{-1}\mathbf{C}_{\mathrm{b}}^{-\mathrm{T}}\mathbf{R}-\mathbf{K}_{\mathrm{Lb}}^{-1}\mathbf{C}_{\mathrm{s}}^{\mathrm{T}}\mathbf{K}_{\mathrm{La}}\left(\mathbf{I}+\mathbf{C}_{\mathrm{s}}\mathbf{K}_{\mathrm{Lb}}^{-1}\mathbf{C}_{\mathrm{s}}^{\mathrm{T}}\mathbf{K}_{\mathrm{La}}\right)^{-1}\mathbf{C}_{\mathrm{s}}\mathbf{K}_{\mathrm{Lb}}^{-1}\mathbf{C}_{\mathrm{b}}^{-\mathrm{T}}\mathbf{R}\right] \tag{36}$$

Comparing Eq. (36) with Eq. (28), the following equation should be true

$$\mathbf{F}_{\mathrm{a}}=\mathbf{K}_{\mathrm{La}}\left(\mathbf{I}+\mathbf{C}_{\mathrm{s}}\mathbf{K}_{\mathrm{Lb}}^{-1}\mathbf{C}_{\mathrm{s}}^{\mathrm{T}}\mathbf{K}_{\mathrm{La}}\right)^{-1}\mathbf{C}_{\mathrm{s}}\mathbf{K}_{\mathrm{Lb}}^{-1}\mathbf{C}_{\mathrm{b}}^{-\mathrm{T}}\mathbf{R} \tag{37}$$

where $\mathbf{I}$ is the identity matrix. In actuality, the aforementioned equation is indeed valid when considered in light of Eq. (22) and the subsequent equation derived by reformulating Eq. (20).

$$\left(\mathbf{C}_{\mathrm{s}}\mathbf{K}_{\mathrm{Lb}}^{-1}\mathbf{C}_{\mathrm{s}}^{\mathrm{T}}\mathbf{K}_{\mathrm{La}}+\mathbf{I}\right)\mathbf{u}_{\mathrm{a}}=\mathbf{C}_{\mathrm{s}}\mathbf{K}_{\mathrm{Lb}}^{-1}\mathbf{C}_{\mathrm{b}}^{-\mathrm{T}}\mathbf{R} \tag{38}$$

It is important to note the distinction between SRI and FDP. The former achieves system reduction by reformulating equilibrium equations based on the concept of pseudo forces, while the latter derives the expression for the inverse of the stiffness matrix by directly applying the given mathematical formulas (SMW formulas). The formulation of Eq. (37) confirms the validity of the system reduction based on the concept of pseudo forces. In other words, the derivation process of SRI clearly reflects the concepts from structural mechanics, and the resulting reduction system possesses a well-defined physical significance. Additionally, the derivation process presented in this paper provides a mechanical explanation for the application of SMW formulas in structural reanalysis.

To determine the number of flops in FDP, Eq. (36) can be reformulated as follows

$$\mathbf{d}=\mathbf{C}_{\mathrm{b}}^{-1}\mathbf{K}_{\mathrm{Lb}}^{-1}\left(\mathbf{P}_{R}-\mathbf{C}_{\mathrm{s}}^{\mathrm{T}}\mathbf{K}_{\mathrm{La}}\mathbf{S}_{\mathrm{a}}\mathbf{C}_{\mathrm{s}}\mathbf{K}_{\mathrm{Lb}}^{-1}\mathbf{P}_{R}\right) \tag{39}$$

where $\mathbf{P}_{R}=\mathbf{C}_{\mathrm{b}}^{-\mathrm{T}}\mathbf{R}$ and $\mathbf{S}_{\mathrm{a}}$ are

$$\mathbf{S}_{\mathrm{a}}=\left(\mathbf{I}+\mathbf{C}_{\mathrm{s}}\mathbf{K}_{\mathrm{Lb}}^{-1}\mathbf{C}_{\mathrm{s}}^{\mathrm{T}}\mathbf{K}_{\mathrm{La}}\right)^{-1} \tag{40}$$

Similar to SRI, $\mathbf{P}_R$, $\mathbf{C}_\mathrm{b}^{-1}$ and $\mathbf{C}_\mathrm{s}$ can be obtained in advance and the calculation amount of obtaining them is not included in computational cost. In other words, the flops of FDP are mainly included in Eqs. (40) and (39). The number of flops for Eqs. (40) and (39) can be determined by

$$\mathbf{S}_\mathrm{a}=\underbrace{\left(\underbrace{\mathbf{I}+\underbrace{\mathbf{C}_\mathrm{s}\overbrace{\mathbf{K}_\mathrm{Lb}^{-1}\overbrace{\mathbf{C}_\mathrm{s}^\mathrm{T}\mathbf{K}_\mathrm{La}}^{nq}}^{2nq}}_{2nq^2+2nq}}_{2nq^2+2nq+q}\right)^{-1}}_{2nq^2+q^3+2nq+q} \tag{41}$$

and

$$\mathbf{d}=\underbrace{\mathbf{C}_\mathrm{b}^{-1}\underbrace{\mathbf{K}_\mathrm{Lb}^{-1}\left(\underbrace{\mathbf{P}_R-\overbrace{\mathbf{C}_\mathrm{s}^\mathrm{T}\underbrace{\mathbf{K}_\mathrm{La}\overbrace{\mathbf{S}_\mathrm{a}\underbrace{\mathbf{C}_\mathrm{s}\overbrace{\mathbf{K}_\mathrm{Lb}^{-1}\mathbf{P}_R}^{n}}_{2nq+n}}^{2q^2+2nq+n}}_{2q^2+2nq+n+q}}^{2q^2+4nq+n+q}}_{2q^2+4nq+2n+q}\right)}_{2q^2+4nq+3n+q}}_{2n^2+2q^2+4nq+3n+q} \tag{42}$$

where the number of flops for the inverse of a $q\times q$ matrix is $q^3$ according to its $O\left(q^3\right)$ complexity. Then, the total number of flops for FDP is

$$T_{F-FDP}=2nq^2+q^3+2n^2+2q^2+6nq+3n+2q \tag{43}$$

*3.3. Comparison study*

This section presents a comparison of the computational cost of SRI, PCG and FDP using Eqs. (32), (33) and (43). Given the set of $n=10000$, the ratios of flops for SRI relative to the other two methods under different settings of $q$ and $k$ are shown in **Fig. 3**. **Fig. 3a**) displays the comparison of flops between SRI and PCG within the range of $q/n\in\left[0.05,0.90\right]$ under the setting of $k_s=0.1,0.3,0.5,0.7$, where $k_s=k/q$ and $k_s=k/n$ for SRI and PCG, respectively. The results indicate that as the ratio of the reduced system size to the total number of DOFs decreases, the computational cost of SRI becomes significantly lower than that of PCG due to the substantial reduction in the size of the equation system to be solved. **Fig. 3b**) presents a comparison of flops between SRI and FDP within the range of $k/q \in [0.01, 0.80]$ under the setting of $q/n = 0.1, 0.3, 0.5, 0.7$. The results demonstrate that as the ratio of the iteration number to the scale of reduced system decreases, the computational cost of SRI can becomes significantly lower than that of FDP. In other words, if an approximate solution can be obtained in a small number of iterations through an effective iterative solution algorithm, SRI has significant advantage in terms of solution efficiency.

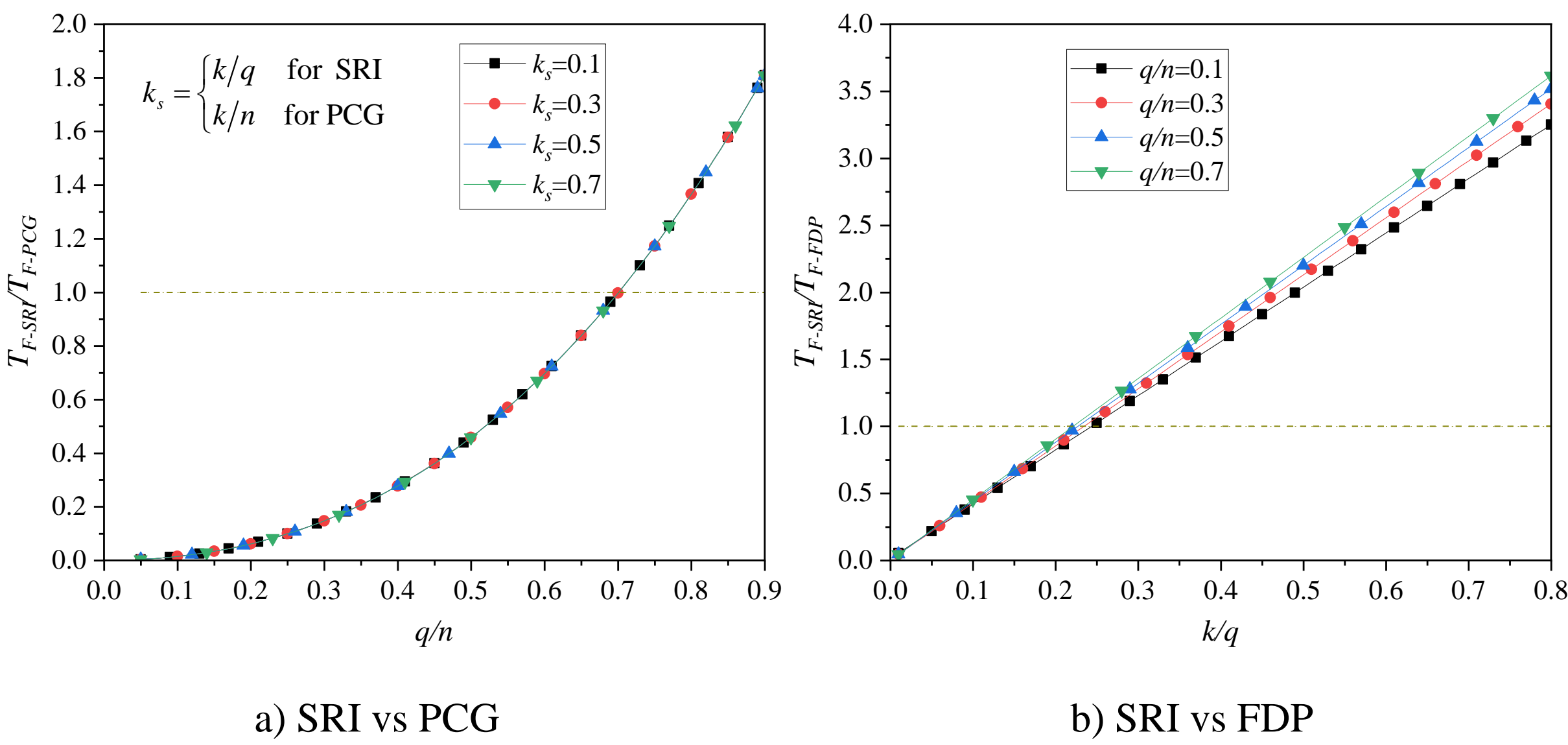

a) SRI vs PCG

b) SRI vs FDP

**Fig. 3**. Comparison of flops for different reanalysis methods.

## 4. Validation

The computational performance of the proposed method is evaluated through numerical examples. The solution algorithms have been programmed in MATLAB and run on a computer having an Intel® Core$^{TM}$ i7-8700 processor and a CPU at 3.2GHz with 64GB of RAM.

### *4.1. Truss structures with linear material*

A two-dimensional truss structure with homogeneous material is adopted to evaluate the computational efficiency of the structural reanalysis methods. The total number of spans and floors is represented by $N_{\mathrm{span}}$ and $N_{\mathrm{floor}}$, respectively. As depicted in **Fig. 4**, both the span and the floor height are set to 5.0m. The horizontal loads applied at the left nodes of the structure are set to $P = 20.0\mathrm{kN}$. The cross-sectional area is set to $2.0\times10^{-3}\mathrm{m}^2$ for each truss element. In the initial structure, Young's modulus is set to $E_0 = 2.0\times10^{11}\,\mathrm{N/m^2}$ for all truss elements. For structural reanalysis, Young's modulus is modified according to the lower and upper values of Young's modulus ($E_{\mathrm{l}}$ and $E_{\mathrm{u}}$). The value of Young's modulus is set gradually from the bottom floor to the top floor, with its lower and upper values of $E_{\mathrm{l}}$ and $E_{\mathrm{u}}$, respectively. for example, in a structure with 5 floors, the values of Young's modulus for the 1st-5th floors are $2.8\times10^{11}\,\mathrm{N/m^2}$, $2.5\times10^{11}\,\mathrm{N/m^2}$, $2.2\times10^{11}\,\mathrm{N/m^2}$, $1.9\times10^{11}\,\mathrm{N/m^2}$, and $1.6\times10^{11}\,\mathrm{N/m^2}$ under the setting of $E_{\mathrm{l}} = 1.6\times10^{11}\,\mathrm{N/m^2}$ and $E_{\mathrm{u}} = 2.8\times10^{11}\,\mathrm{N/m^2}$, respectively. For a structure with $N_{\mathrm{span}}$ spans, the diagonal bars in the 2nd - $N_{\mathrm{span}}$ th spans (Dotted lines) are defined as additional components. Thus, the number of additional components $N_{\mathrm{m}}$ and the ratio of $N_{\mathrm{m}}$ to the total number of DOFs are determined based on the setting of $N_{\mathrm{span}}$. The tolerance for iterative solution algorithms is set to $1\times10^{-12}$.

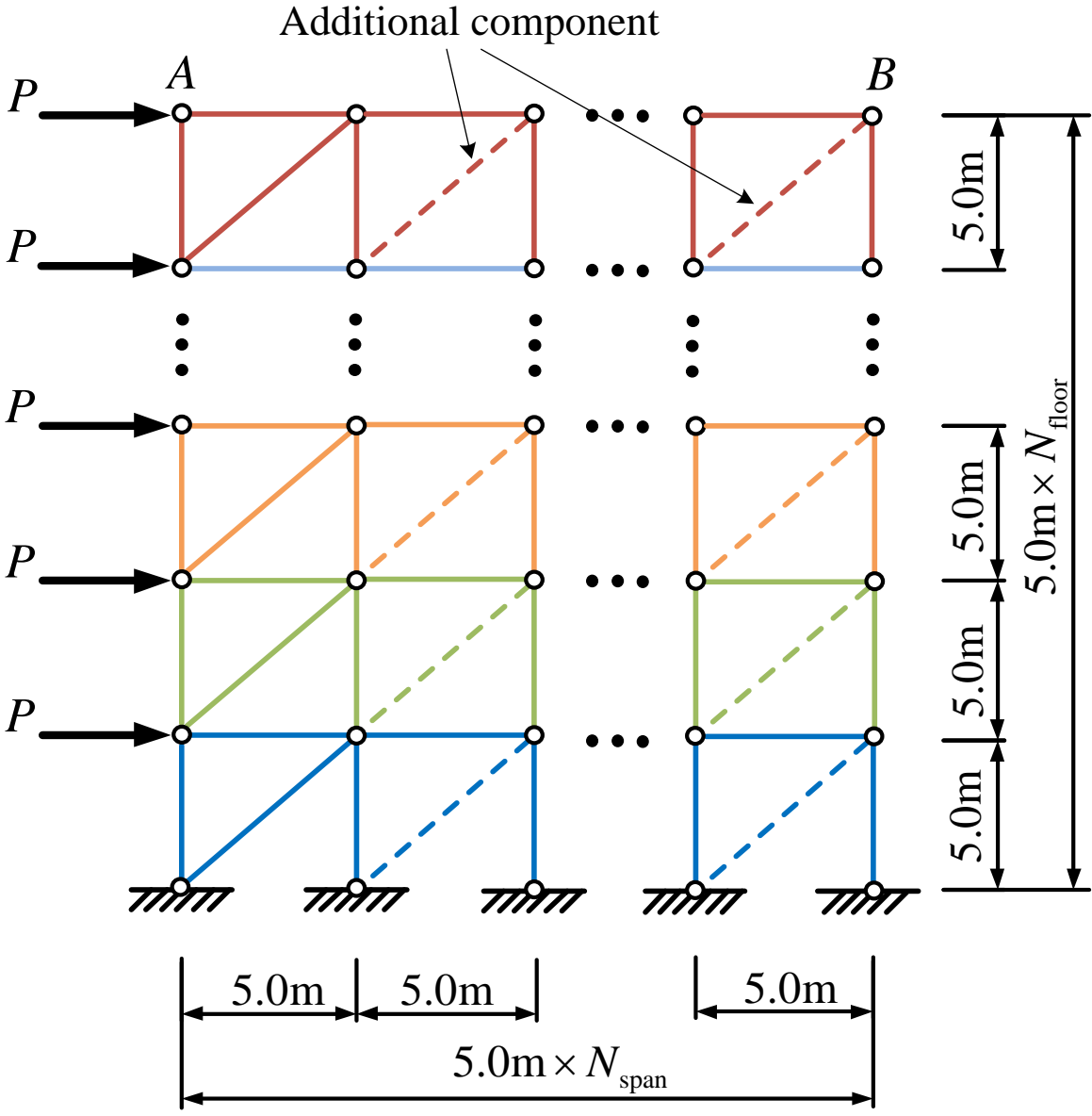

**Fig. 4**. The two-dimensional truss with $N_{\mathrm{span}}$ spans and $N_{\mathrm{floor}}$ floors.

A parameter $a$ is used to determine the number of spans, floors, and the ratio of the number of additional components to the total number of DOFs, denoted by $N_{\text{m}}/N_{\text{DOF}}$. The relationships between $N_{\text{span}}$, $N_{\text{floor}}$, and $a$ are expressed as

$$2^a - 1 = N_{\text{span}}, N_{\text{floor}} \times \left(N_{\text{span}} + 1\right) = N_{\text{node}} \tag{44}$$

where $N_{\text{node}}$ represents the total number of unconstrained nodes of the structure. Numerical models of varying scales are considered, with three settings of $N_{\text{node}}$ ($N_{\text{node}}$ = 2048, 4096 and 6144). The corresponding values are presented in **Table 2**.

**Table 2** Setting of number of spans and floors according to parameter $a$.

| $a$ | $N_{\text{span}}$ | $N_{\text{floor}}$ | | | $N_{\text{m}}/N_{\text{DOF}}$ |
|---|---|---|---|---|---|
| | | $N_{\text{node}}$ = 2048 | $N_{\text{node}}$ = 4096 | $N_{\text{node}}$ = 6144 | |
| 1 | 1 | 1024 | 2048 | 3072 | 0.000 |
| 2 | 3 | 512 | 1024 | 1536 | 0.250 |
| 3 | 7 | 256 | 512 | 768 | 0.375 |
| 4 | 15 | 128 | 256 | 384 | 0.438 |
| 5 | 31 | 64 | 128 | 192 | 0.469 |
| 6 | 63 | 32 | 64 | 96 | 0.484 |

Under the setting of $E_{\text{l}} = 0.5 \times 10^{11}\ \text{N/m}^2$ and $E_{\text{u}} = 3.5 \times 10^{11}\ \text{N/m}^2$, the displacement solutions of the structure are obtained using FDP, PCG, SRI and the conventional method (regular complete analysis). With three scales of numerical models ($N_{\text{node}}$ = 2048, 4096 and 6144), the displacements of nodes $A$ and $B$ under the setting of $N_{\text{span}} = 31$ are presented in **Table 3**. The results demonstrate that the displacement solutions obtained by the four methods are consistent, verifying the solution accuracy of the proposed method.

The Relative Computational Time (RCT) is defined as follows to evaluate the computational efficiency of the reanalysis methods.

$$\text{RCT} = T_{\text{reanalysis}}/T_{\text{c}} \tag{45}$$

where $T_{\text{reanalysis}}$ and $T_{\text{c}}$ represent the computational time of the structural reanalysis method and the conventional method, respectively. The lower and upper values of Young's modulus are set to $E_l = 0.5\times10^{11}\ \mathrm{N/m^2}$ and $E_u = 3.5\times10^{11}\ \mathrm{N/m^2}$, reaching as much as 80% of the modification range of Young's modulus. For the cases of $N_{node} = 2048$, $N_{node} = 4096$ and $N_{node} = 6144$, the total number of structural DOFs are 4096, 8192, and 12288, respectively; the average computational time of the conventional method is 1.3374s, 9.5132s and 27.9936s, respectively. For the cases of $N_{node} = 2048$ and $N_{node} = 6144$, the relatively computational time of FDP, PCG and SRI under different settings of parameter $a$ is presented in **Fig. 5**. The computational efficiency of these four structural reanalysis methods is summarized as follows.

**Table 3** Displacements of the truss with $N_{span} = 31$, $E_l = 0.5\times10^{11}\ \mathrm{N/m^2}$ and $E_u = 3.5\times10^{11}\ \mathrm{N/m^2}$.

| $N_{node}$ | Method | Node *A* | | Node *B* | |
|---|---|---|---|---|---|
| | | $d_h$ /m | $d_v$ /m | $d_h$ /m | $d_v$ /m |
| 2048 | FDP | $2.327843\times10^{-1}$ | $3.694581\times10^{-2}$ | $2.117298\times10^{-1}$ | $-6.198756\times10^{-2}$ |
| | PCG | $2.327843\times10^{-1}$ | $3.694581\times10^{-2}$ | $2.117298\times10^{-1}$ | $-6.198756\times10^{-2}$ |
| | SRI | $2.327843\times10^{-1}$ | $3.694581\times10^{-2}$ | $2.117298\times10^{-1}$ | $-6.198756\times10^{-2}$ |
| | Conventional | $2.327843\times10^{-1}$ | $3.694581\times10^{-2}$ | $2.117298\times10^{-1}$ | $-6.198756\times10^{-2}$ |
| 4096 | FDP | $2.485152\times10^{0}$ | $3.272211\times10^{-1}$ | $2.462131\times10^{0}$ | $-4.393270\times10^{-1}$ |
| | PCG | $2.485152\times10^{0}$ | $3.272211\times10^{-1}$ | $2.462131\times10^{0}$ | $-4.393270\times10^{-1}$ |
| | SRI | $2.485152\times10^{0}$ | $3.272211\times10^{-1}$ | $2.462131\times10^{0}$ | $-4.393270\times10^{-1}$ |
| | Conventional | $2.485152\times10^{0}$ | $3.272211\times10^{-1}$ | $2.462131\times10^{0}$ | $-4.393270\times10^{-1}$ |
| 6144 | FDP | $1.167079\times10^{1}$ | $1.161943\times10^{0}$ | $1.164704\times10^{1}$ | $-1.418954\times10^{0}$ |
| | PCG | $1.167079\times10^{1}$ | $1.161943\times10^{0}$ | $1.164704\times10^{1}$ | $-1.418954\times10^{0}$ |
| | SRI | $1.167079\times10^{1}$ | $1.161943\times10^{0}$ | $1.164704\times10^{1}$ | $-1.418954\times10^{0}$ |
| | Conventional | $1.167079\times10^{1}$ | $1.161943\times10^{0}$ | $1.164704\times10^{1}$ | $-1.418954\times10^{0}$ |

Note: $d_h$ and $d_v$ represent the horizontal displacement and vertical displacement, respectively.

FDP exhibits high computational efficiency when $N_m/N_{DOF}$ is relatively small, such as $N_m/N_{DOF} < 0.25$. With the increase in $N_m/N_{DOF}$, the RCT of FDP significantly increases, while the solution efficiency remarkably decreases. This is because FDP still needs to solve large-scale linear equations using the direction method (such as the full factorization method) in these cases, leading to the increased computational time. For PCG, the computational efficiency is consistently higher than that of the conventional method and there is no significant relationship between the value of RCT and the value of $N_m/N_{DOF}$. The results in **Fig. 5** demonstrate that the computational efficiency of PCG is improved for large-scale structures. If the value of $N_{\mathrm{m}}/N_{\mathrm{DOF}}$ is high and the scale of structural numerical models is large, the computational efficiency of PCG is higher than that of FDP.

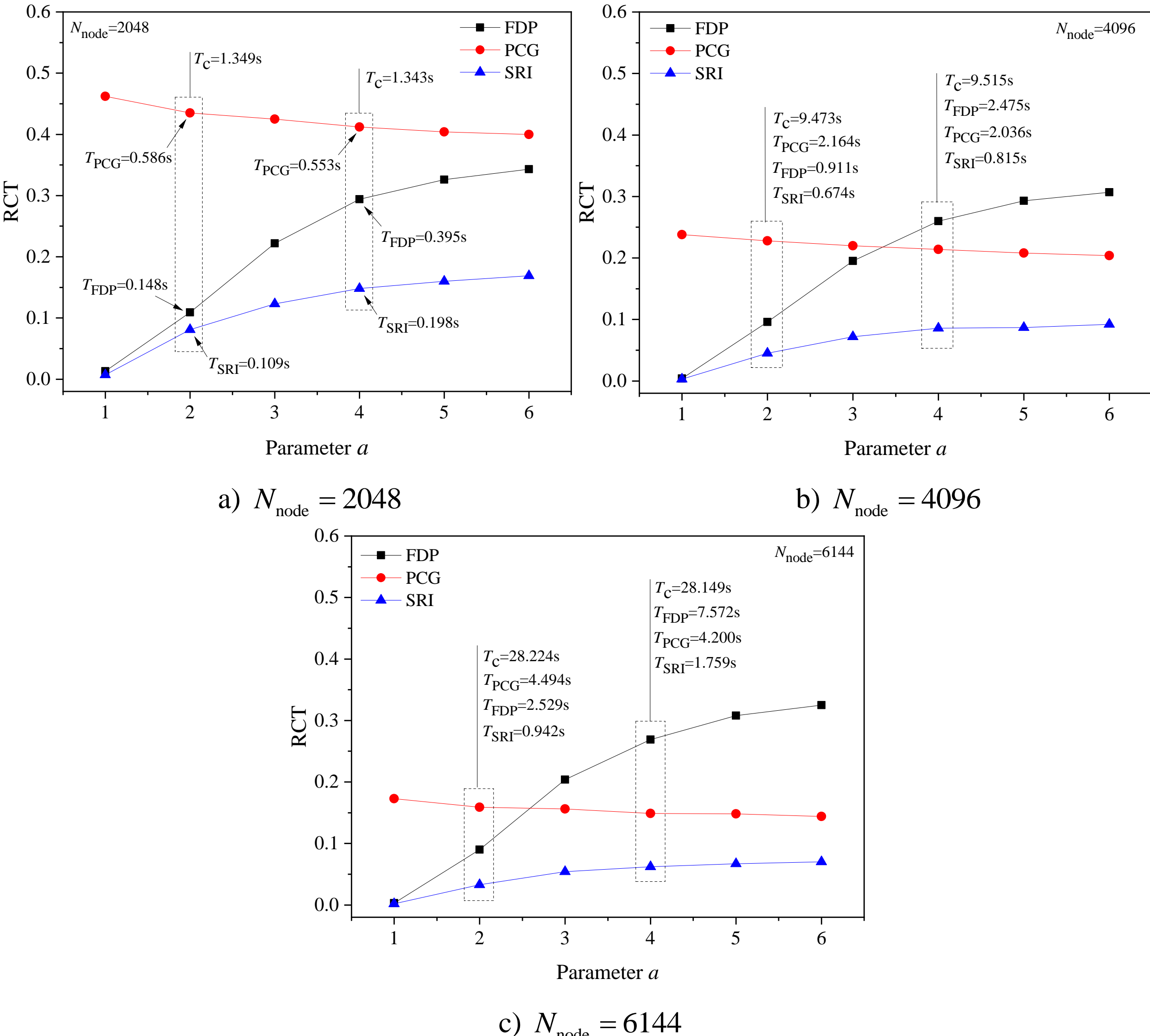

a) $N_{\mathrm{node}} = 2048$ b) $N_{\mathrm{node}} = 4096$

c) $N_{\mathrm{node}} = 6144$

**Fig. 5**. RCT of the three reanalysis methods under different settings of parameter *a*.

Compared to FDP, SRI maintains high solution efficiency for the cases with a high value of $N_{\mathrm{m}}/N_{\mathrm{DOF}}$. The reason is that the pre-conditioned iterative solution method used in SRI can significantly reduce the computational time required to solve the large-scale linear equations. Compared to PCG, SRI only needs to solve a linear system on a relatively small scale, with fewer calculations. In other words, SRI can leverage the advantages of system reduction and pre-conditioned iterative solution technology while maintaining high computational efficiency under different conditions. Additionally, the RCT of SRI significantly decreases with the increase in the scale of the numerical model.

### *4.2. Frame structures with linear material*

A two-dimensional frame structure (**Fig. 6**) is employed to demonstrate the effectiveness of the proposed method in the structural reanalysis of frame structures. As shown in **Fig. 6**, both the span and the floor height are set to 5.0m. The horizontal loads applied at the left nodes of the structure are set to $P = 20.0\text{kN}$. The cross-section for each beam or column is set to $b \times h = 0.10\text{m} \times 0.3\text{m}$. The structure is simulated using the beam finite element based on Euler-Bernoulli beam theory, with $N_{sb}$ denoting the number of elements used to model each beam. **Fig. 6** provides the definition of the basis system and additional components for the frame structure with $N_{sb} = 2$, where the first beam element in each beam (the red element) is defined as the additional component.

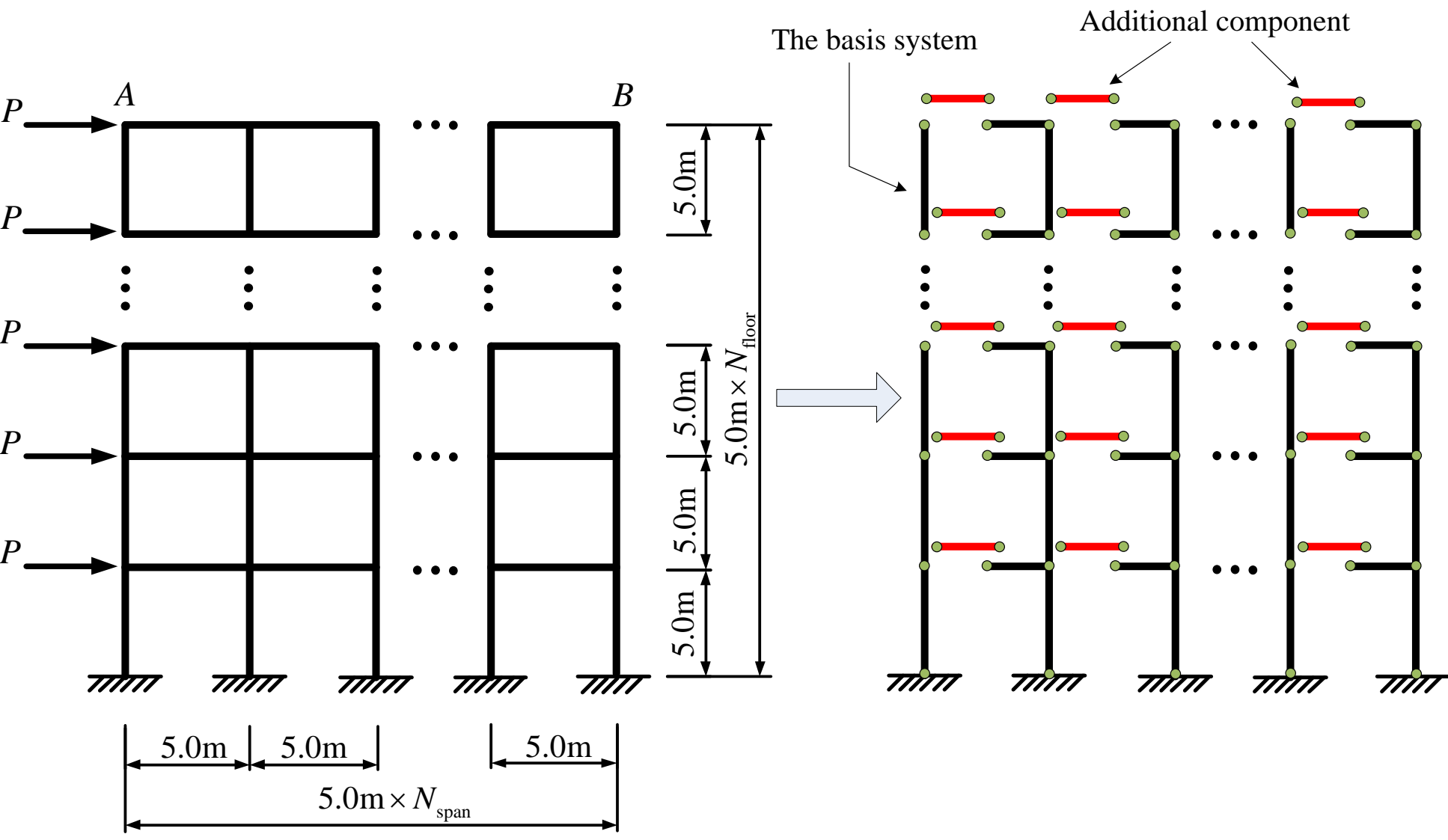

**Fig. 6**. The two-dimensional frame and its basis system and additional components.

(1) Homogeneous material

For beams and columns with homogeneous material, the cross-sectional area and moment of inertia are set to $3.0\times10^{-2}\,\mathrm{m}^2$ and $2.25\times10^{-4}\,\mathrm{m}^4$, respectively. For the initial structure, Young's modulus is set to $E_0 = 2.0\times10^{11}\,\mathrm{N/m^2}$ for all elements. In the structural reanalysis, Young's modulus is adjusted in accordance with the lower and upper values of Young's modulus ($E_\mathrm{l}$ and $E_\mathrm{u}$). The value of Young's modulus is incrementally assigned from the bottom floor to the top floor, with its lower and upper values being $E_\mathrm{l}$ and $E_\mathrm{u}$, respectively, analogous to the example presented in **Sec. 4.1**. Under varying settings of $N_\mathrm{sb}$, the total number of nodes $N_\mathrm{node}$ and the ratio of the number of additional components to the number of nodes, denoted by $N_\mathrm{m}/N_\mathrm{node}$, for the structure with $N_\mathrm{span} = 50$, are listed in **Table 4**.

Regarding the implementation of reanalysis using FDP and SRI, the stiffness parameter matrix of the *e*-th element $\mathbf{k}_{\mathrm{L}(e)}$ is obtained as

$$\mathbf{k}_{\mathrm{L}(e)} = \frac{1}{L_e^3}\begin{bmatrix} 2E_eA_eL_e^2 & 0 & 0 \\ 0 & 2E_eI_eL_e^2 & 0 \\ 0 & 0 & 6E_eI_e\left(L_e^2+4\right) \end{bmatrix} \tag{46}$$

where $E_e$, $A_e$, $I_e$, and $L_e$ represent Young's modulus, cross-sectional area, the moment of inertia of the cross-section, and the length of the *e*-th element, respectively.

**Table 4** Number of nodes and ratio of additional components under different settings of $N_\mathrm{sb}$.

| $N_\mathrm{sb}$ | $N_\mathrm{node}$ | | | | $N_\mathrm{m}/N_\mathrm{node}$ |
|---|---|---|---|---|---|
| | $N_\mathrm{floor} = 20$ | $N_\mathrm{floor} = 30$ | $N_\mathrm{floor} = 40$ | $N_\mathrm{floor} = 50$ | |
| 1 | 1020 | 1530 | 2040 | 2550 | 0.980 |
| 2 | 2020 | 3030 | 4040 | 5050 | 0.495 |
| 3 | 3020 | 4530 | 6040 | 7550 | 0.331 |
| 4 | 4020 | 6030 | 8040 | 10050 | 0.249 |

Firstly, the solution precision of these reanalysis methods is examined for the structure with $N_\mathrm{span} = 50$ and $N_\mathrm{floor} = 20$, utilizing the displacement solution of the conventional method (regular complete analysis) as a benchmark. Under $E_\mathrm{l} = 0.4\times10^{11}\,\mathrm{N/m^2}$ and $E_\mathrm{u} = 3.6\times10^{11}\,\mathrm{N/m^2}$, the displacement solutions at node *B* under different values of $N_{\text{sb}}$ are presented in **Table 5**, wherein the tolerance for PCG and SRI is set to $1\times10^{-12}$. The results indicate that the proposed method can yield accurate displacement solutions, as evidenced by the congruent solutions attained by the four methods.

**Table 5** Displacements at node *B* of the frame structure with $N_{\text{span}}=50$ and $N_{\text{floor}}=20$.

| $N_{\text{sb}}$ | Methods | $d_h$ /m | $d_v$ /m | Rotation/rad |
|---|---|---|---|---|
| 1 | FDP | $3.444080\times10^{-2}$ | $-3.476257\times10^{-4}$ | $-1.044827\times10^{-4}$ |
| | PCG | $3.444080\times10^{-2}$ | $-3.476257\times10^{-4}$ | $-1.044827\times10^{-4}$ |
| | SRI | $3.444080\times10^{-2}$ | $-3.476257\times10^{-4}$ | $-1.044827\times10^{-4}$ |
| | Conventional | $3.444080\times10^{-2}$ | $-3.476257\times10^{-4}$ | $-1.044827\times10^{-4}$ |
| 2 | FDP | $3.444080\times10^{-2}$ | $-3.476257\times10^{-4}$ | $-1.044827\times10^{-4}$ |
| | PCG | $3.444080\times10^{-2}$ | $-3.476257\times10^{-4}$ | $-1.044827\times10^{-4}$ |
| | SRI | $3.444080\times10^{-2}$ | $-3.476257\times10^{-4}$ | $-1.044827\times10^{-4}$ |
| | Conventional | $3.444080\times10^{-2}$ | $-3.476257\times10^{-4}$ | $-1.044827\times10^{-4}$ |
| 3 | FDP | $3.444080\times10^{-2}$ | $-3.476257\times10^{-4}$ | $-1.044827\times10^{-4}$ |
| | PCG | $3.444080\times10^{-2}$ | $-3.476257\times10^{-4}$ | $-1.044827\times10^{-4}$ |
| | SRI | $3.444080\times10^{-2}$ | $-3.476257\times10^{-4}$ | $-1.044827\times10^{-4}$ |
| | Conventional | $3.444080\times10^{-2}$ | $-3.476257\times10^{-4}$ | $-1.044827\times10^{-4}$ |
| 4 | FDP | $3.444080\times10^{-2}$ | $-3.476257\times10^{-4}$ | $-1.044827\times10^{-4}$ |
| | PCG | $3.444080\times10^{-2}$ | $-3.476257\times10^{-4}$ | $-1.044827\times10^{-4}$ |
| | SRI | $3.444080\times10^{-2}$ | $-3.476257\times10^{-4}$ | $-1.044827\times10^{-4}$ |
| | Conventional | $3.444080\times10^{-2}$ | $-3.476257\times10^{-4}$ | $-1.044827\times10^{-4}$ |

Note: $d_h$ and $d_v$ represent the horizontal displacement and vertical displacement, respectively.

Furthermore, the solution efficiency of FDP, PCG, and SRI is appraised. With varying settings of $N_{\text{sb}}$ and $N_{\text{floor}}$, the computational time of the conventional method and the relative computational time of FDP, PCG, and SRI are presented in **Table 6**. The results in **Table 6** are expounded as follows. In instances of $N_{\text{sb}}=1$ and $N_{\text{sb}}=2$, the scale of the reduction system reconstructed based on the additional components is relatively large, making the advantages of system reduction in FDP and SRI unable to be fully reflected. Consequently, PCG achieves the highest computational efficiency in these cases. The computational efficiency of FDP and SRI is markedly enhanced when the scale of the reduction system reconstructed based on the additional components is relatively small, such as $N_{\text{sb}}=3$ and $N_{\text{sb}}=4$. The solution efficiency of FDP may be inferior to that of PCG (such as the instance of $N_{\text{sb}}=3$ and $N_{\text{floor}}>20$) with an increase in the aggregate number of DOFs.

This is attributable to the necessity for a considerable amount of computation to solve the large-scale reduction system via the direct solution method.

**Table 6** Comparison of solution efficiency between different methods ( $N_{\text{span}} = 50$ ).

| $N_{\text{sb}}$ ( $N_{\text{m}}/N_{\text{node}}$ ) | $N_{\text{floor}}$ | $N_{\text{DOF}}$ | $T_{\text{c}}$ /s | RCT | | |
|---|---|---|---|---|---|---|
| | | | | FDP | PCG | SRI |
| 1 (0.980) | 10 | 1530 | 0.059 | 2.515 | 0.785 | 2.601 |
| | 20 | 3060 | 0.373 | 2.335 | 0.641 | 1.744 |
| | 30 | 4590 | 1.306 | 2.198 | 0.441 | 1.089 |
| | 40 | 6120 | 2.879 | 2.175 | 0.365 | 0.863 |
| | 50 | 7650 | 5.470 | 2.156 | 0.303 | 0.705 |
| 2 (0.495) | 10 | 3030 | 0.370 | 0.585 | 0.578 | 0.744 |
| | 20 | 6060 | 2.853 | 0.478 | 0.329 | 0.385 |
| | 30 | 9090 | 9.023 | 0.466 | 0.244 | 0.265 |
| | 40 | 12120 | 21.143 | 0.432 | 0.194 | 0.199 |
| | 50 | 15150 | 40.509 | 0.419 | 0.157 | 0.161 |
| 3 (0.331) | 10 | 4530 | 1.168 | 0.238 | 0.405 | 0.339 |
| | 20 | 9060 | 9.166 | 0.192 | 0.228 | 0.170 |
| | 30 | 13590 | 29.985 | 0.181 | 0.165 | 0.114 |
| | 40 | 18120 | 69.152 | 0.175 | 0.131 | 0.085 |
| | 50 | 22650 | 139.444 | 0.167 | 0.103 | 0.068 |
| 4 (0.249) | 10 | 6030 | 2.868 | 0.123 | 0.292 | 0.186 |
| | 20 | 12060 | 21.604 | 0.110 | 0.173 | 0.095 |
| | 30 | 18090 | 71.509 | 0.104 | 0.126 | 0.064 |
| | 40 | 24120 | 156.277 | 0.102 | 0.106 | 0.049 |
| | 50 | 30150 | 332.567 | 0.090 | 0.102 | 0.038 |

In contrast to FDP, SRI circumvents the issue of diminished computational efficiency resulting from an increase in the scale of the reduced system by employing a pre-conditioned iterative method to solve the reduced system. Consequently, SRI retains high computational efficiency even for large-scale numerical models. As indicated in **Table 6**, the advantage of SRI in terms of computational efficiency becomes increasingly pronounced as the proportion of additional components decreases and the scale of the numerical model increases.

(2) Functionally graded material

The utilization of functionally graded (FG) structures in aerospace, marine, civil construction is expanding rapidly due to their high strength-to-weight ratio [39]. To ensure the accuracy of numerical simulations, a refined mesh is typically required for modeling FG structures using finite element method. This significantly increases the computational burden of finite element solutions and presents considerable challenges for structural optimization and reliability assessment that require repeated finite element solutions. Consequently, fast reanalysis of modified FG structures has become a valuable endeavor that provides important support for efficient optimization design and reliability assessment of FG structures.

The effective Young's modulus of the FG beams is assumed to vary continuously through the beam height according to a power-law distribution as follows [40, 41]:

$$E(y)=\left(E_{US}-E_{LS}\right)\left(\frac{y}{h}+\frac{1}{2}\right)^{p}+E_{LS},\quad -\frac{h}{2}\leq y\leq\frac{h}{2} \tag{47}$$

where $E(y)$ represents the Young's modulus of the FG material at location $y$ through the height, $E_{US}$ and $E_{LS}$ denote the generic material properties at the upper and lower surfaces of the FG beam, respectively; $h$ refers to the height of the FG beam, $p$ is the power-law exponent.

A variety of different beam models have been established to conduct analysis of FG beams, such as Euler-Bernoulli beam model [41, 42], first-order shear beam model [40, 43, 44] and higher-order shear beam model [39, 45-47]. In this study, the Euler Bernoulli beam model is selected and the stiffness matrix of the *i*-th FG beam element in local coordinate system is expressed as [41, 42]

$$\mathbf{k}_{c(i)}=\frac{1}{L_i^3}\begin{bmatrix} A_{Ei}b_iL_i^2 & 0 & -B_{Ei}b_iL_i^2 & -A_{Ei}b_iL_i^2 & 0 & B_{Ei}b_iL_i^2 \\ 0 & 12D_{Ei}b_i & 6D_{Ei}b_iL_i & 0 & -12D_{Ei}b_i & 6D_{Ei}b_iL_i \\ -B_{Ei}b_iL_i^2 & 6D_{Ei}b_iL_i & 4D_{Ei}b_iL_i^2 & B_{Ei}b_iL_i^2 & -6D_{Ei}b_iL_i & 2D_{Ei}b_iL_i^2 \\ -A_{Ei}b_iL_i^2 & 0 & B_{Ei}b_iL_i^2 & A_{Ei}b_iL_i^2 & 0 & -B_{Ei}b_iL_i^2 \\ 0 & -12D_{Ei}b_i & -6D_{Ei}b_iL_i & 0 & 12D_{Ei}b_i & -6D_{Ei}b_iL_i \\ B_{Ei}b_iL_i^2 & 6D_{Ei}b_iL_i & 2D_{Ei}b_iL_i^2 & -B_{Ei}b_iL_i^2 & -6D_{Ei}b_iL_i & 4D_{Ei}b_iL_i^2 \end{bmatrix} \tag{48}$$

where

$$A_{Ei}=\frac{h_i}{p_i+1}E_{US}^{i}+\frac{h_i p_i}{p_i+1}E_{LS}^{i} \tag{49}$$

$$B_{Ei}=\frac{h_i^2}{2\left(p_i+1\right)\left(p_i+2\right)}E_{US}^{i}-\frac{h_i^2}{2\left(p_i+1\right)\left(p_i+2\right)}E_{LS}^{i} \tag{50}$$

$$D_{Ei}=\frac{h_i^3\left(p_i^2+p_i+2\right)}{4\left(p_i+1\right)\left(p_i+2\right)\left(p_i+3\right)}E_{US}^{i}+\left[\frac{h_i^3}{12}-\frac{h_i^3\left(p_i^2+p_i+2\right)}{4\left(p_i+1\right)\left(p_i+2\right)\left(p_i+3\right)}\right]E_{LS}^{i} \tag{51}$$

With the transform matrix between the stiffness matrix and stiffness parameter matrix in local coordinate system [34] used, the element stiffness parameter matrix for the FG beam element can be obtained as

$$\mathbf{k}_{\mathrm{L}(i)}=\mathbf{c}_{\mathrm{c}(i)}\mathbf{k}_{c(i)}\mathbf{c}_{\mathrm{c}(i)}^{\mathrm{T}}=\begin{bmatrix} k_{11}^{(i)} & k_{12}^{(i)} & 0 \\ k_{12}^{(i)} & k_{22}^{(i)} & 0 \\ 0 & 0 & k_{33}^{(i)} \end{bmatrix}=\begin{bmatrix} 2A_{Ei}b_i/L_i & -2B_{Ei}b_i/L_i & 0 \\ -2B_{Ei}b_i/L_i & 2D_{Ei}b_i/L_i & 0 \\ 0 & 0 & 6\left(L_i^2+4\right)D_{Ei}b_i/L_i^3 \end{bmatrix} \tag{52}$$

where $\mathbf{c}_{\mathrm{c}(i)}$ represents the matrix of $\mathbf{c}_{(i)}$ corresponding to the element local coordinate system.

It can be realized that $k_{12}^{(i)}$ reflects the coupling effect of axial deformation and flexural deformation. The comparison between Eq. (46) and Eq. (52) indicates that, different from the element with homogeneous material, $\mathbf{k}_{\mathrm{L}(i)}$ of the FG beam element is no longer a diagonal matrix, due to the coupling of element stiffness parameters.

The two-dimensional frame structure shown in **Fig. 6** is taken into consideration, with the columns and beams of the frame assumed to be composed of FG material and the effective Young's modulus varying continuously through the height of the cross-section. Specifically, $E_{LS}$ and $E_{US}$ represent the values of Young's modulus at the upper and lower surfaces of the beams, respectively, or at the right and left surfaces of the columns, respectively. For the initial structure, $E_{LS}$ and $E_{US}$ are set to the same value as $E_{LS}=E_{US}=E_0=2.0\times10^{11}\,\mathrm{N/m^2}$ and the power-law exponent is set as $p=1.0$ for all elements. For reanalysis, the structure is modified by changing the value of $E_{US}\in\left[E_{\mathrm{l}},E_{\mathrm{u}}\right]$. The value of $E_{US}$ is set gradually from the bottom floor to the top floor, with its lower and upper values being $E_{\mathrm{l}}$ and $E_{\mathrm{u}}$, respectively. In numerical modeling, each column or beam is modelled using 8 FG beam elements so as to obtain acceptable approximate displacement solutions, based on the investigation provided by Li et al. (See Table 3 in Ref. [39]).

Using the displacement solution of the conventional method (regular complete analysis) as a reference, the solution accuracy of these reanalysis methods is evaluated for the frame structure with $N_{\mathrm{span}}=N_{\mathrm{floor}}=4$. Under $E_{\mathrm{l}}=0.4\times10^{11}\,\mathrm{N/m^2}$ and $E_{\mathrm{u}}=3.6\times10^{11}\,\mathrm{N/m^2}$, the displacement solutions at node $B$ for different settings of the power-law exponent are shown in **Table 7**, where the tolerance for PCG and SRI is set to $1\times10^{-12}$. The results demonstrate that the proposed method can yield accurate displacement solutions, indicating that the expressions of stiffness parameters for the functionally graded beam element derived using the transformation matrix obtained from the spectral decomposition of the homogeneous material beam element are reliable.

**Table 7** Displacements at node *B* of the frame with $N_{\text{span}} = N_{\text{floor}} = 4$.

| $p$ | Methods | $d_h$ /m | $d_v$ /m | Rotation/rad |
|---|---|---|---|---|
| 0.5 | FDP | $2.111726\times10^{-2}$ | $-5.859733\times10^{-4}$ | $-8.018189\times10^{-4}$ |
| | PCG | $2.111726\times10^{-2}$ | $-5.859733\times10^{-4}$ | $-8.018189\times10^{-4}$ |
| | SRI | $2.111726\times10^{-2}$ | $-5.859733\times10^{-4}$ | $-8.018189\times10^{-4}$ |
| | Conventional | $2.111726\times10^{-2}$ | $-5.859733\times10^{-4}$ | $-8.018189\times10^{-4}$ |
| 1.0 | FDP | $1.757305\times10^{-2}$ | $-7.509972\times10^{-4}$ | $-3.540270\times10^{-4}$ |
| | PCG | $1.757305\times10^{-2}$ | $-7.509972\times10^{-4}$ | $-3.540270\times10^{-4}$ |
| | SRI | $1.757305\times10^{-2}$ | $-7.509972\times10^{-4}$ | $-3.540270\times10^{-4}$ |
| | Conventional | $1.757305\times10^{-2}$ | $-7.509972\times10^{-4}$ | $-3.540270\times10^{-4}$ |
| 2.0 | FDP | $1.717977\times10^{-2}$ | $-6.925002\times10^{-4}$ | $-2.863039\times10^{-4}$ |
| | PCG | $1.717977\times10^{-2}$ | $-6.925002\times10^{-4}$ | $-2.863039\times10^{-4}$ |
| | SRI | $1.717977\times10^{-2}$ | $-6.925002\times10^{-4}$ | $-2.863039\times10^{-4}$ |
| | Conventional | $1.717977\times10^{-2}$ | $-6.925002\times10^{-4}$ | $-2.863039\times10^{-4}$ |

Note: $d_h$ and $d_v$ represent the horizontal displacement and vertical displacement, respectively.

The solution efficiency of FDP, PCG, SRI and the conventional method is investigated for the structures with $N_{\text{span}} = 10$. With different settings of $N_{\text{floor}}$ and the power-law exponent $p$, the computational time of the different methods is presented in **Table 8**. It is worth noting that the ratio of the number of additional components to the total number of DOFs is $N_{\text{m}}/N_{\text{DOF}} = 0.063$ for all cases in **Table 8**. In this scenario, both FDP and SRI can significantly reduce the computational time, as evidenced by the results in the table. Data from **Table 8** reveals that SRI has the highest solution efficiency, indicating that the proposed reanalysis method performs well for modifications of statically indeterminate structures composed of functionally graded beams. Further statistics on the results in **Table 8** show that, the relative computational time of SRI defined by $T_{\text{SRI}}/T_{\text{c}}$ is about $4.89\%$, $3.46\%$, $2.34\%$ and $1.88\%$, respectively, for the four structural scales where the total number of structural DOFs is 4740, 9480, 14220 and 18960, respectively. This suggests that, with the increase of the structural scale, the solution efficiency of SRI is improved more significantly.

**Table 8** Comparison of solution efficiency between different methods ( $N_{\text{span}} = 10$ ).

| $p$ | $N_{\text{floor}}$ | $N_{\text{DOF}}$ | $T_{\text{c}}$ /s | $T_{\text{FDP}}$ /s | $T_{\text{PCG}}$ /s | $T_{\text{SRI}}$ /s |
|---|---|---|---|---|---|---|
| 0.5 | 10 | 4740 | 1.420 | 0.157 | 0.761 | 0.069 |
| | 20 | 9480 | 11.034 | 0.931 | 3.510 | 0.379 |
| | 30 | 14220 | 35.562 | 2.826 | 8.204 | 0.842 |
| | 40 | 18960 | 83.397 | 6.665 | 14.856 | 1.649 |
| 1.0 | 10 | 4740 | 1.446 | 0.153 | 0.517 | 0.067 |
| | 20 | 9480 | 11.411 | 0.964 | 2.107 | 0.379 |
| | 30 | 14220 | 36.822 | 2.954 | 4.725 | 0.824 |
| | 40 | 18960 | 85.889 | 6.827 | 8.568 | 1.474 |
| 2.0 | 10 | 4740 | 1.415 | 0.143 | 0.388 | 0.075 |
| | 20 | 9480 | 11.115 | 0.937 | 1.575 | 0.395 |
| | 30 | 14220 | 35.717 | 2.913 | 3.617 | 0.842 |
| | 40 | 18960 | 83.944 | 6.706 | 6.479 | 1.613 |
| 5.0 | 10 | 4740 | 1.431 | 0.155 | 0.326 | 0.068 |
| | 20 | 9480 | 11.145 | 0.924 | 1.320 | 0.391 |
| | 30 | 14220 | 35.887 | 2.823 | 2.964 | 0.888 |
| | 40 | 18960 | 83.704 | 6.669 | 5.214 | 1.565 |
| 10.0 | 10 | 4740 | 1.571 | 0.157 | 0.283 | 0.077 |
| | 20 | 9480 | 11.334 | 0.971 | 1.177 | 0.394 |
| | 30 | 14220 | 37.463 | 2.816 | 2.561 | 0.857 |
| | 40 | 18960 | 83.732 | 6.385 | 4.477 | 1.588 |

*4.3. Truss structures with nonlinear material*

In this section, the two-dimensional truss structure shown in **Fig. 4** is assumed to be composed of nonlinear material to assess the effectiveness of the proposed reanalysis method for structural static nonlinear analysis. The bilinear material model employed in Ref. [22] is used, where $E_0$ and $E_{\text{t}}$ represent Young's modulus and the tangent modulus, respectively, and $\sigma_y$ refers to the yield stress. Young's modulus and the tangent modulus are set to $E_0 = 2.0 \times 10^{11}\,\text{N/m}^2$ and $E_{\text{t}} = 0.3 \times 10^{11}\,\text{N/m}^2$, respectively. The cross-sectional area of all elements is set to $2.0 \times 10^{-2}\,\text{m}^2$.

For static nonlinear analysis, the equilibrium equations of the structure are expressed as

$$\mathbf{F}(\mathbf{d}) - \lambda_p \mathbf{P}_0 = \mathbf{0} \tag{53}$$

where $\mathbf{P}_0$ represents the constantly applied force vector, and $\lambda_p$ indicates the load factor. The Newton-Raphson method with load control [48] is employed to solve the above nonlinear equation.

In implementing of the Newton-Raphson method, the following incremental equations are required to be repeatedly solved.

$$\mathbf{K}_{\mathrm{t}}\big|_{\mathbf{d}=\mathbf{d}^*}\Delta\mathbf{d} = -\mathbf{R}_{\mathrm{r}}\left(\mathbf{d}^*\right) \tag{54}$$

where $\mathbf{K}_{\mathrm{t}}\big|_{\mathbf{d}=\mathbf{d}^*}$ indicates the tangent stiffness matrix of the structure and $\mathbf{R}_{\mathrm{r}}\left(\mathbf{d}\right) = \mathbf{F}\left(\mathbf{d}\right) - \lambda_p \mathbf{P}_0$ is the residual force vector of the structure. In this work, the proposed reanalysis method is employed to solve the above linearized equation.

Three methods are employed to obtain the incremental displacements for implementation of the incremental-iteration procedure: (1) NR-Regular, where the incremental displacements are obtained by solving Eq. (54) with the full factorization method used; (2) NR-Reduction, where the incremental displacements are obtained by solving Eq. (24) with the full factorization method used; (3) NR-SRI, where the incremental displacements are obtained using the proposed reanalysis method. The convergence criterion for iteration in the Newton-Raphson method is set to $\left\|\mathbf{R}_{\mathrm{r}}\right\| / \left\|\lambda_p \mathbf{P}_0\right\| < 1\times10^{-8}$ . For NR-SRI, the convergence criterion for iteration is set to $\left\|\mathbf{r}_j\right\| / \left\|\lambda_p \mathbf{P}_0\right\| < 1\times10^{-15}$.

Let $N_{\mathrm{span}} = 30$ and $N_{\mathrm{floor}} = 150$, the number of DOFs and finite element of the entire numerical model is 9300 and 13650, respectively. The basic value of the load is set to $P_0 = 5\times10^4\,\mathrm{N}$ and 20 load steps with equal incremental load ( $2.5\times10^3\,\mathrm{N}$ ) are adopted to obtain the displacement solutions of the structure. In three yield stress cases of $\sigma_y = 4.5\times10^7\ \mathrm{N/m^2}$ , $\sigma_y = 2.5\times10^7\ \mathrm{N/m^2}$ , and $\sigma_y = 0.5\times10^7\ \mathrm{N/m^2}$ , the relation curves between the horizontal displacement of node *B* and the load factor obtained by the Newton-Raphson method using the three methods to obtain the incremental displacements are illustrated in **Fig. 7**. The results revealed that the displacement solutions obtained by NR-Regular, NR-Reduction, and NR-SRI are consistent. Thus, the proposed reanalysis method can be effectively integrated into the solution algorithm for structural nonlinear analysis without compromising solution accuracy.

**Table 9** illustrates the computational time required by NR-Regular, NR-Reduction, and NR-SRI, as well as the total number of nonlinear elements under the three settings of yield stress. **Table** 9 demonstrates that the nonlinear analysis algorithm incorporating the proposed reanalysis method has the highest solution efficiency, with its computational time being approximately 10% of that required by NR-Regular and 28% of that required by NR-Reduction. Compared to NR-Regular, the improvement in computational efficiency achieved by NR-Reduction is derived from the reduction in the scale of linear equations to be solved during the nonlinear analysis process. In contrast to NR-Reduction, the improvement in computational efficiency achieved by NR-SRI reflects the high efficiency of using pre-conditioned conjugate gradient method to solve linear equations. In cases with different numbers of nonlinear elements, NR-SRI consistently exhibits high computational performance, verifying that the reanalysis method proposed in this paper can play a positive role in the static nonlinear analysis of structures.

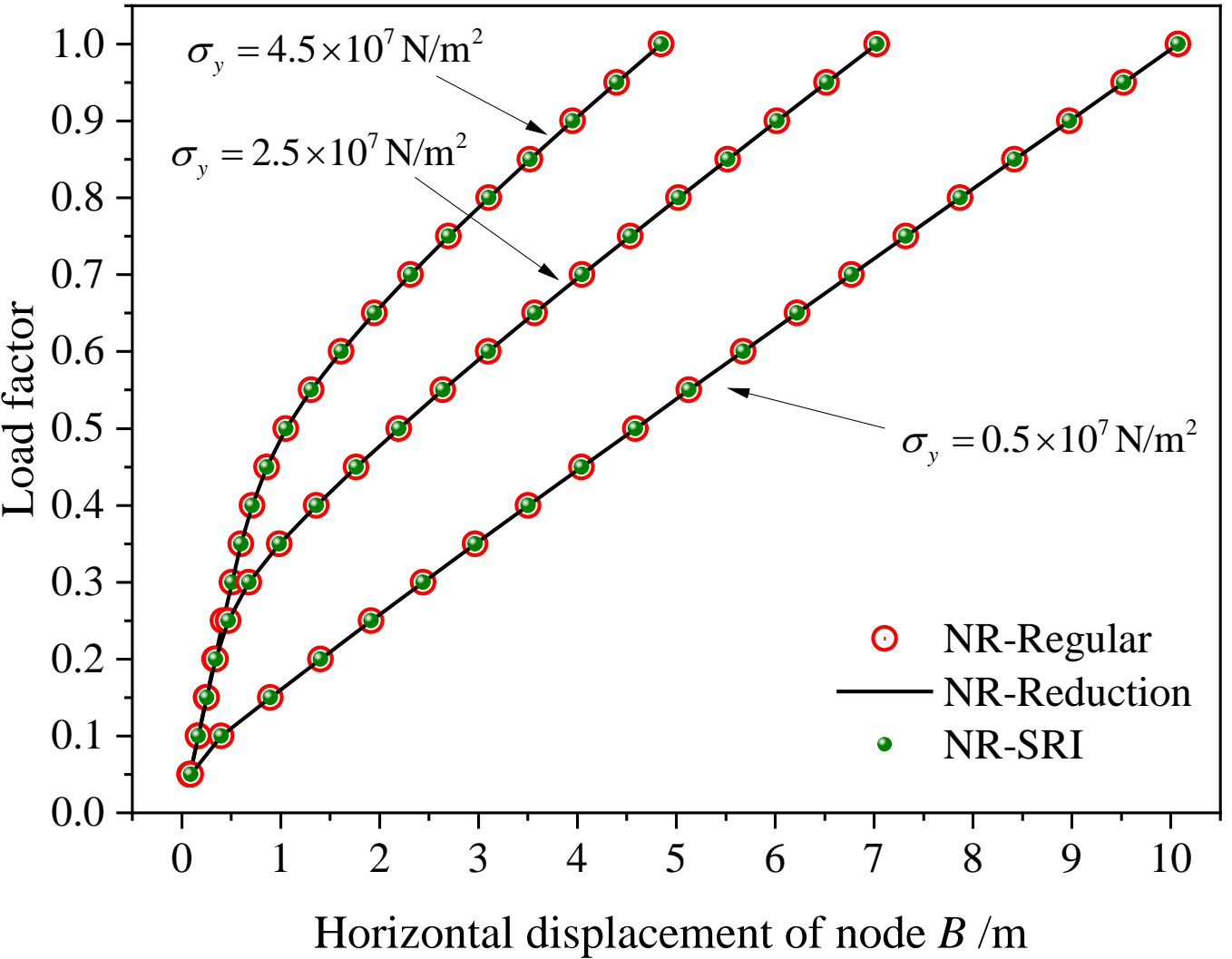

**Fig. 7**. Relation curves between displacement and load factor.

**Table 9** Comparison of solution efficiency for NR-Regular, NR-Reduction and NR-SRI.

| $\sigma_y$ /$10^7$ N/m$^2$ | $N_{\text{NLE}}$ | $T_{\text{NR-Regular}}$ /s | $T_{\text{NR-Reduction}}$ /s | $T_{\text{NR-SRI}}$ /s |
|---|---|---|---|---|
| 4.5 | 1691 | 689.610 | 225.257 | 61.203 |
| 2.5 | 2567 | 794.655 | 265.741 | 72.963 |
| 0.5 | 9116 | 937.395 | 321.751 | 95.491 |

Note: $N_{\text{NLE}}$ represents the total number of nonlinear elements, $T_{\text{NR-Regular}}$ , $T_{\text{NR-Reduction}}$ and $T_{\text{NR-SRI}}$ are the total computational time for NR-Regular, NR-Reduction and NR-SRI, respectively.

## 5. Conclusions

In this paper, a system reduction-based approximate reanalysis method is developed for statically indeterminate structures with high-rank modification. The computational cost is examined and numerical examples are provided. The conclusions are drawn as follows.

(1) Through the separation of the basis system and additional components and the introduction of spectral decomposition, the reduced equation system of statically indeterminate structures can be successfully established, providing support for the implementation of a fast approximate reanalysis method. Unlike the flexibility disassembly perturbation method, which is based on the given mathematical formulas (SMW formulas), the present derivation has clear mechanical significance and provides a mechanical explanation for the application of SMW formulas in structural reanalysis.

(2) The proposed reanalysis method realizes the integration of system reduction and iterative solution. Numerical results indicate that this method exhibits high computational performance for statically indeterminate structures with high-rank modification due to its advantages in system reduction and fast iterative solution. Generally, the advantage in solution efficiency of the proposed method becomes increasingly pronounced with the increase in the scale of the numerical model. The combination of system reduction and pre-conditioned iterative solution can be considered as a promising mode for developing high-performance reanalysis methods.

(3) For the statically indeterminate structures composed of functionally graded material, where multiple beam elements are required to model each column and beam, the proposed system reduction method can significantly reduce the equation system to be solved, thereby enabling the proposed reanalysis method to achieve excellent computational performance. Results from numerical examples indicate that the proposed method maintains high computational performance even in cases with coupled element stiffness parameters.

(4) The proposed reanalysis method can be applied to improve the performance of static nonlinear analysis for statically indeterminate structures with nonlinear materials.

The proposed reanalysis method has been successfully applied to structures with high-rank modification, while the following limitations should be addressed to make it more applicable in the future. Firstly, this method requires an invariant mesh definition. In cases where the mesh is variable, repeated pre-processing is necessary, increasing computational workload and reducing reanalysis efficiency. Secondly, the nonlinear analysis method incorporating the proposed reanalysis algorithm is not suitable for solving problems with geometric nonlinearity because system reduction is based on fixed structural configuration and topological connectivity. Lastly, this reanalysis method is not suitable for cases where the number of additional components is relatively large compared to the number of structural DOFs because the advantage of system reduction in this method is not significant.

## Acknowledgments

The project is funded by the National Natural Science Foundation of China (Grant No. 52178209, Grant No. 51878299) and Guangdong Basic and Applied Basic Research Foundation, China (Grant No. 2021A1515012280, Grant No. 2020A1515010611).

## Nomenclature

| | |
|---|---|
| $\mathbf{c}_{(i)}$, $\mathbf{C}_{(i)}$ | The transform matrices of the *i*-th element reflected on the sequence of element DOFs and structural DOFs |
| $\mathbf{C}$ | The transform matrix between structural stiffness parameter matrix and structural stiffness matrix |
| $\mathbf{k}_{(i)}$, $\mathbf{k}_{\mathrm{L}(i)}$ | The stiffness matrix and stiffness parameter matrix of the *i*-th element |
| $\mathbf{K}$, $\mathbf{K}_{\mathrm{L}}$ | The stiffness matrix and stiffness parameter matrix of the structure |
| $\mathbf{d}$ | The nodal displacement vector of the structure |
| $\mathbf{R}$ | The external load vector of the structure |
| BASESYS | The set of elements in the basis system |

ADDSYS The set of elements in the additional components

$\mathbf{K}_{\mathrm{b}}$, $\mathbf{K}_{\mathrm{a}}$ The stiffness matrices for the basis system and the additional components

$\mathbf{C}_{\mathrm{b}}$, $\mathbf{C}_{\mathrm{a}}$ The transform matrices for the basis system and the additional components

$\mathbf{K}_{\mathrm{Lb}}$, $\mathbf{K}_{\mathrm{La}}$ The stiffness parameter matrices for the basis system and the additional components

$\mathbf{u}_{(i)}$, $\mathbf{u}_{\mathrm{a}}$ The generalized deformation vectors of the *i*-th element and for the additional components

$\mathbf{F}_{\mathrm{a}}$ The generalized force vector for the additional components

$E_0$, $E_{\mathrm{l}}$, $E_{\mathrm{u}}$ The initial Young's modulus, the lower and upper values of Young's modulus

$E(y)$ The Young's modulus of FG material at location *y* through the height

$\mathbf{K}_{\mathrm{t}}\big|_{\mathbf{d}=\mathbf{d}^*}$ The tangent stiffness matrix of the structure

$\mathbf{R}_{\mathrm{r}}(\mathbf{d})$ The residual force vector of the structure

FDP The structural reanalysis method based on Flexibility Disassembly Perturbation

PCG The existing structural reanalysis method based on the Pre-conditioning Conjugate Gradient method

SRI The structural reanalysis method based on System Reduction and Iterative solution